\documentclass[twocolumn, trackchanges]{aastex631}
\usepackage{subfigure}
\usepackage{multirow}

\newcommand\CO{$^{12}$CO }
\newcommand\COl{$^{13}$CO }
\newcommand\COll{C$^{18}$O }

\submitjournal{ApJ}
\shorttitle{VSF analysis of Galactic molecular clouds}
\shortauthors{Ma et al.}

\begin{document}
\defcitealias{Kainulainen2009}{K09}
\defcitealias{Kolmogorov1941a}{K41}
\defcitealias{She1994}{SL94}
\defcitealias{Boldyrev2002a}{B02}
\title{Examining Turbulence in Galactic Molecular Clouds - I: A Statistical Analysis of Velocity Structures}

\correspondingauthor{Yuehui Ma, Xuepeng Chen, Hongchi Wang}
\email{mayh@pmo.ac.cn, xpchen@pmo.ac.cn, hcwang@pmo.ac.cn}
\author[0000-0002-8051-5228]{Yuehui Ma}
\affil{Purple Mountain Observatory and Key Laboratory of Radio Astronomy, Chinese Academy of Sciences, 10 Yuanhua Road, Nanjing 210033, China}

\author[0000-0002-6388-649X]{Miaomiao Zhang}
\affil{Purple Mountain Observatory and Key Laboratory of Radio Astronomy, Chinese Academy of Sciences, 10 Yuanhua Road, Nanjing 210033, China}

\author[0000-0003-0746-7968]{Hongchi Wang}
\affil{Purple Mountain Observatory and Key Laboratory of Radio Astronomy, Chinese Academy of Sciences, 10 Yuanhua Road, Nanjing 210033, China}
\affil{School of Astronomy and Space Science, University of Science and Technology of China, Hefei, Anhui 230026, China}

\author[0000-0001-8060-1321]{Min Fang}
\affil{Purple Mountain Observatory and Key Laboratory of Radio Astronomy, Chinese Academy of Sciences, 10 Yuanhua Road, Nanjing 210033, China}
\affil{School of Astronomy and Space Science, University of Science and Technology of China, Hefei, Anhui 230026, China}

\author[0009-0009-3431-1150]{Zhenyi Yue}
\affil{Purple Mountain Observatory and Key Laboratory of Radio Astronomy, Chinese Academy of Sciences, 10 Yuanhua Road, Nanjing 210033, China}
\affil{School of Astronomy and Space Science, University of Science and Technology of China, Hefei, Anhui 230026, China}

\author[0000-0003-3151-8964]{Xuepeng Chen}
\affil{Purple Mountain Observatory and Key Laboratory of Radio Astronomy, Chinese Academy of Sciences, 10 Yuanhua Road, Nanjing 210033, China}
\affil{School of Astronomy and Space Science, University of Science and Technology of China, Hefei, Anhui 230026, China}

\author[0000-0001-7768-7320]{Ji Yang}
\affil{Purple Mountain Observatory and Key Laboratory of Radio Astronomy, Chinese Academy of Sciences, 10 Yuanhua Road, Nanjing 210033, China}

\author[0000-0002-7489-0179]{Fujun Du}
\affil{Purple Mountain Observatory and Key Laboratory of Radio Astronomy, Chinese Academy of Sciences, 10 Yuanhua Road, Nanjing 210033, China}

\author[0000-0002-0197-470X]{Yang Su}
\affil{Purple Mountain Observatory and Key Laboratory of Radio Astronomy, Chinese Academy of Sciences, 10 Yuanhua Road, Nanjing 210033, China}
\affil{School of Astronomy and Space Science, University of Science and Technology of China, Hefei, Anhui 230026, China}

\author[0009-0004-2947-4020]{Suziye He}
\affil{Purple Mountain Observatory and Key Laboratory of Radio Astronomy, Chinese Academy of Sciences, 10 Yuanhua Road, Nanjing 210033, China}
\affil{School of Astronomy and Space Science, University of Science and Technology of China, Hefei, Anhui 230026, China}

\author[0000-0003-1714-0600]{Haoran Feng}
\affil{Purple Mountain Observatory and Key Laboratory of Radio Astronomy, Chinese Academy of Sciences, 10 Yuanhua Road, Nanjing 210033, China}
\affil{School of Astronomy and Space Science, University of Science and Technology of China, Hefei, Anhui 230026, China}

\author[0000-0002-3904-1622]{Yan Sun}
\affil{Purple Mountain Observatory and Key Laboratory of Radio Astronomy, Chinese Academy of Sciences, 10 Yuanhua Road, Nanjing 210033, China}
\affil{School of Astronomy and Space Science, University of Science and Technology of China, Hefei, Anhui 230026, China}

\author[0000-0003-2218-3437]{Chong Li}
\affil{Purple Mountain Observatory and Key Laboratory of Radio Astronomy, Chinese Academy of Sciences, 10 Yuanhua Road, Nanjing 210033, China}

\author[0000-0003-4586-7751]{Qing-Zeng Yan}
\affil{Purple Mountain Observatory and Key Laboratory of Radio Astronomy, Chinese Academy of Sciences, 10 Yuanhua Road, Nanjing 210033, China}

\author[0000-0003-0849-0692]{Zhiwei Chen}
\affil{Purple Mountain Observatory and Key Laboratory of Radio Astronomy, Chinese Academy of Sciences, 10 Yuanhua Road, Nanjing 210033, China}
\affil{Center for Astronomy and Space Sciences, Three Gorges University, Yichang 443002, China}

\author[0000-0003-2549-7247]{Shaobo Zhang}
\affil{Purple Mountain Observatory and Key Laboratory of Radio Astronomy, Chinese Academy of Sciences, 10 Yuanhua Road, Nanjing 210033, China}

\author[0000-0003-2418-3350]{Xin Zhou}
\affil{Purple Mountain Observatory and Key Laboratory of Radio Astronomy, Chinese Academy of Sciences, 10 Yuanhua Road, Nanjing 210033, China}

\begin{abstract}
    We present a systematic analysis of the velocity structure functions (VSFs) of 167 molecular clouds with angular sizes greater than $\sim$176 arcmin$^2$ in three sectors of the Galactic mid-plane. We calculated the 1st- to 3rd-order VSFs and found that 60\% of the VSFs exhibit power-law distributions. The relative power-law exponents are consistent with predictions from intermittent turbulence models. Column density weighting reduces the proportion of power-law VSFs and steepens the VSF slopes, implying a reduction of turbulent energy in high-density regions. All clouds show small-scale intermittency, with slightly stronger intermittency in those molecular clouds showing none power-law VSFs. Negative VSF exponents that may indicate gravitational collapse are not observed in our sample. The scaling exponents of the observed VSFs do not correlate with the virial parameters of the molecular clouds. These two observations suggest that gravity-dominated scales in molecular clouds still need further investigation. Consistent VSF scaling exponents for the molecular clouds with significant power-law VSFs suggest large-scale external driving of turbulence in these molecular clouds. However, the driving mechanisms are likely not universal, as the power-law scaling coefficients in our results show relatively large scatter. The fact that nearly 40\% of the VSFs deviate to some extent from power-law distributions suggests that the influence of local environments on the internal turbulence of molecular clouds may not be negligible.
\end{abstract}


\section{Introduction} \label{sec1}
Turbulence is a common but complex phenomenon in fluid dynamics, characterized by chaotic and irregular fluid motion \citep{Frisch1995}. In the interstellar medium (ISM), turbulence is inherently multi-phase and multi-scale. It plays an essential role in the formation of molecular clouds and their internal substructures such as filaments, clumps, and cores \citep{Banerjee2009, Moeckel2015}. In the framework of the turbulence support (gravo-turbulence, GT) model that describes the structure of molecular clouds, turbulent motions can counteract self-gravity of molecular clouds, providing support against collapse on relatively large scales ($\sim$pc) \citep{Vazquez-Semadeni2003, Padoan2011}. The shock waves induced by supersonic turbulence within molecular clouds can create dense regions that further collapse to form stars. Consequently, turbulence is considered an essential physical process to regulate the star formation rate/efficiency within molecular clouds \citep{Mac2004, Ballesteros-Paredes2007}. However, there is a debate on the effectiveness of this GT regime since the observed large line widths of molecular clouds might not represent supersonic turbulence but anisotropic gravitational collapse, according to the global hierarchical collapse (GHC) model \citep{Vazquez-Semadeni2019, Vazquez-Semadeni2024}. From the perspective of understanding the formation of molecular clouds and stars, a series of questions remain to be answered. For example, whether molecular clouds are turbulent and within what scale range turbulence is dominant. Given that turbulent energy dissipates \citep{MacLow1998}, therefore, if molecular clouds are turbulence-dominated, what is the predominant energy source?
%In contrast, gravitational and magnetic instabilities seem to be responsible for creating larger-scale structures, like cloud complexes and superclouds \citep{Vazquez-Semadeni1995, Dobbs2014}. 

The dynamics of a turbulent flow are believed to be described by the Navier-Stokes equation, to which a general analytic solution is intractable because of the existence of a non-linear item in the equation. Although chaotic, the spatial or temporal averaged properties of turbulent flows can be investigated statistically using mathematical methods such as probability distribution function (PDF), high-order PDF statistical moments, structure function (SF), power spectrum (PS), $\Delta$-variance algorithm based on wavelet analysis, and principle component analysis (PCA) \citep{Gill1990, Vazquez-Semadeni1994, Heyer1997, Pope2001, Ossenkopf2008a, Ossenkopf2008b, Burkhart2021}. Among the above methods, PDFs and SFs are the two most straightforward. \cite{Ma2021, Ma2022} investigated the column density PDFs (N-PDFs) of molecular clouds in two sectors of the outer Galaxy and found that the molecular clouds are mainly dominated by turbulence ($\sim$70$\%$ showing log-normal N-PDFs) regardless of their distances, sizes, and masses, at least on cloud scales ($\sim$pc). However, PDF is a one-dimensional description of the turbulent density field that strongly depends on an ideal tracer of molecular gas accounting for a broad enough dynamic range of H$_2$ column density and the ability to resolve projection problems \citep{Goodman2009}. Comparatively, the velocity structure function (VSF) is a two-point statistical tool that measures the $p$-th order moments of the velocity differences at a given spatial separation \citep{Kolmogorov1941a}. It directly provides insights into the scaling properties of turbulence and helps in understanding the distribution and intensity of turbulent fluctuations over different scales. 

Mathematically, the $p$-th order VSF is defined as the following formula
\begin{equation}
    \label{eq1}
    S_p(\mathbf{l}) = \langle |\mathbf{v}_{\mathbf{x}} - \mathbf{v}_{\mathbf{x+l}}|^p \rangle,
\end{equation}
where $\mathbf{v}$, $\mathbf{x}$, and $\mathbf{l}$ are the velocity, position, and spatial lag vectors, respectively. In practice, after azimuthally averaging across all vector angles, these vectors can be expressed as scalars. The VSFs are expected to be power-laws, $S(p) =\nu_0 l^{\zeta_p}$, within the so-called ``inertial range'' \citep{Frisch1995}. 

\cite{Kolmogorov1941a, Kolmogorov1941b} (hereafter \citetalias{Kolmogorov1941a}) made the first quantitative prediction, $\zeta_p = p/3$, about the energy distribution in turbulent flows, based on Richardson's cascade theory of turbulence \citep{Richardson1922}. \citetalias{Kolmogorov1941a}'s predictions rely on three assumptions, i.e., very high Reynolds number (universality), small-scale isotropy, and an inertial range where energy transfer depends only on average transfer rate. However, extensive experimental and numerical studies have shown that the \citetalias{Kolmogorov1941a} theory is not entirely accurate, with results revealing notable deviations from $\zeta_p=p/3$ for $p>3$ \citep{Batchelor1949, Anselmet1984, Vincen1991}. This deviation is attributed to intermittency, referring to the irregular occurrence of intense and localized bursts of rapid velocity changes within the turbulent flow. Several inertial-range intermittency models have been proposed to refine the \citetalias{Kolmogorov1941a} theory, such as the log-normal \citep{Kolmogorov1962}, fractal \citep{Frisch1978}, multifractal \citep{Schertzer1987, Meneveau1991}, log-Poisson (\citealp{She1994}, hereafter \citetalias{She1994}) models, and those based on vortex geometry \citep{Jimenez1998}. Among these models, the \citetalias{She1994} model is highly influential. It proposes that turbulence can be characterized by a hierarchy of fluctuation events with increasing intensities related through hierarchical symmetry. The model provides a formula for the scaling exponent of the $p$-th order VSF for incompressible turbulence, which is 
\begin{equation}
    \label{eq2}
    \zeta_p = \gamma p + C(1-\beta^{p})  = \frac{p}{9} + 2[1-(\frac{2}{3})^{\frac{p}{3}}],
\end{equation}
where $C = 2$ is the co-dimension of the most intermittent structure of the flow (vortex filaments with a dimension of one in \citetalias{She1994}'s model), and $\beta$ is a parameter that measures the degree of intermittency. In the limit of $\beta \rightarrow 1$, there is no intermittency, while in the limit of $\beta \rightarrow 0$, only the most intermittent structures persist in the flow \citep{She1995, She2009}. Here, in Eq. \ref{eq2}, $\beta^3 = 2/3$. The parameter $\gamma$ is related to $C$ and $\beta$, and it measures the singularity index of the most intermittent structure \citep{She1995}. 

The \citetalias{She1994} model has been validated by extensive experimental and numerical data, and different scaling law formulas have been developed based on \citetalias{She1994} theory \citep{She2009}. As an extension of the \citetalias{She1994} theory, \cite{Boldyrev2002a} (hereafter \citetalias{Boldyrev2002a}) proposed a model for supersonic turbulence, primarily to describe turbulence in the ISM and star formation processes with the assumption that the most dissipative structures in turbulence are sheet-like shocks. The \citetalias{Boldyrev2002a} formula of the VSF scaling exponent can be expressed as follows  
\begin{equation}
    \label{eq3}
    \zeta_p = \gamma p + C(1-\beta^{p})  = \frac{p}{9} + [1-(\frac{1}{3})^{\frac{p}{3}}],
\end{equation}
with $\gamma = 1/9$, $C=$1, and $\beta^3 = 1/3$. 

With theoretical predictions on the shape and the power law exponents of the VSFs, we can examine whether the velocity fields of molecular clouds conform to the features of turbulent motions. The existence of power-laws in VSFs and the corresponding power-law exponents have been examined in various numerical simulations in hydrodynamic or magnetohydrodynamic situations \citep{Boldyrev2002b, Padoan2004, Joung2006, Kritsuk2007, Schmidt2009, Federrath2010}. The results from \cite{Schmidt2009} and \cite{Federrath2010} have shown that the scaling exponent of the VSFs is strongly affected by the turbulence driving mode. \cite{Kritsuk2007} found that for compressible turbulence, the Kolmogorov scaling of the power spectra and structure functions can be recovered by calculating VSFs using the density-weighted velocity. \cite{Chira2019} found in their simulations that the exponents of VSFs are sensitive to physical processes like accretion and supernova explosion. In contrast to numerical simulations, there are relatively fewer observational studies concerning VSFs of molecular clouds. Power-law VSFs are obtained within the spatial range from sub-pc to a few tens of parsecs in some local molecular clouds, and the corresponding velocity fields contain small-scale ($\sim$0.02 pc) filamentary intermittent structures that align with local magnetic fields \citep{Pety2003, Padoan2003, Hily-Blant2008}. \citet{Heyer2004} examined VSFs of 27 Galactic molecular clouds derived via the PCA approach and found that the VSFs exhibit a narrow range of scaling exponents and coefficients, emphasizing consistent turbulent behavior across different molecular clouds. Following \citet{Heyer2004}, \cite{Roman-Duval2011} applied the same PCA method to the Galactic Ring Survey data, and found that the VSF exponents are consistent with both Burgers and compressible intermittent turbulence. Although the conclusions of \cite{Heyer2004} and \cite{Roman-Duval2011} were drawn from sample studies, with the former using 27 clouds and the latter 367, both studies used the PCA method instead of directly examining the moments of velocity increments. The scaling distributions of the VSFs of Galactic molecular clouds still need to be investigated using large-scale survey data.

In this work, we conduct a sample analysis of the VSFs of molecular clouds in three sectors of the first, second, and third quadrants of the Galactic mid-plane, using the $^{12}$CO and $^{13}$CO J=1-0 emission line data from the Milky Way Imaging Scroll Painting (MWISP) survey \citep{Su2019}. The paper is organized as follows. In Section \ref{sec2}, we briefly introduce the observations, the molecular cloud samples used in this work, and the methods for calculating VSFs. Section \ref{sec3} presents the statistics of the obtained VSFs, including discussions on the existence of power-law distributions and whether the power-law exponents are consistent with turbulence theoretical predictions. We discuss the results and the implications for future investigations in Section \ref{sec4} and summarize our results in Section \ref{sec5}.

\section{Data and Methods} \label{sec2}
\subsection{Observations and Data} \label{sec2.1}
In this work, we use the \COl data from the MWISP survey to calculate the centroid velocity maps of the molecular clouds. The current coverage of the MWISP survey is from $l = 12^{\circ}$ to $l = 230^{\circ}$ within $b=\pm5 \arcdeg$. One can find an earlier detailed introduction about the observations, data reduction processes, and data quality of the MWISP survey by \cite{Su2019}. Here we briefly give some basic information about the survey and data. For the MWISP project, the \CO, \COl, and \COll $J=1-0$ spectra are obtained simultaneously using the nine-beam Superconducting Spectroscopic Array Receiver (SSAR) \citep{Shan2012} equipped on the PMO-13.7 m telescope. The surveyed area is divided into individual units of sizes of 30$\arcmin\times$30$\arcmin$ during the observation. Eventually, a processed FITS file of the same size will be created for each surveyed unit. The half-power beam width (HPBW) of the PMO-13.7 m telescope is around 52$\arcsec$ and 50$\arcsec$ at 110 GHz and 115 GHz, respectively, while the pixel size of the final data is 30$\arcsec\times$30$\arcsec$ along the directions of Galactic longitude and latitude. The spatial resolution of the data can provide us with details of molecular clouds at a physical resolution of $\sim$0.25 pc at a distance of 1 kpc. The spectrometer of the PMO-13.7 m telescope has 16,384 velocity channels and a total bandwidth of 1 GHz, which provides a velocity resolution of 0.17 km s$^{-1}$ at 110 GHz. The sensitivity of the MWISP data is $\sim$0.5 K per channel in the \CO $J=1-0$ line and $\sim$0.3 K per channel in the \COl $J=1-0$ and \COll $J=1-0$ lines. The statistical properties of the noise of MWISP data are given in \cite{Cai2021}.

\subsection{Sample of Molecular Clouds}\label{sec2.2}
In observational data, we define molecular clouds as structures composed of contiguous position-position-velocity (PPV) voxels above a given intensity threshold. In this work, we used the molecular clouds identified by \cite{Yan2021} using the MWISP \CO data. \cite{Yan2021} have identified 2214 molecular clouds in three regions of the Galactic plane, i.e. $l = [25.\!\!\arcdeg8, 49.\!\!\arcdeg7], [104.\!\!\arcdeg75, 150.\!\!\arcdeg25], [209.\!\!\arcdeg75, 219.\!\!\arcdeg75]$ and $b \le \pm5\arcdeg$, from the MWISP \CO data using a clustering algorithm named the Density-Based Spatial Clustering of Applications with Noise (DBSCAN) \citep{Ester1996}. In the first quadrant, \cite{Yan2021} focused on nearby molecular clouds within a velocity range of [$-$5, 30] km s$^{-1}$, partially avoiding velocity confusion caused by overlapping spiral arms. In the selected second and third quadrant regions, the spiral arms are more clearly separated in velocity space, so the molecular clouds are not affected by velocity crowding. The velocity ranges are [-95, 25] km s$^{-1}$ and [0, 70] km s$^{-1}$ for molecular clouds in the second and third quadrant areas, respectively. However, as is well-known, \CO is an optically thick tracer of molecular gas. Compared to \CO, \COl emission is relatively optically thin and exhibits fewer pseudo multiple velocity components caused by self-absorption. Moreover, \COl is also a better tracer for the column density fields of molecular clouds than \CO. We then implemented the DBSCAN algorithm to the \COl data within the PPV volumes defined by the boundaries of the \CO clouds identified by \cite{Yan2021}, meaning that we assigned each molecular cloud with corresponding \COl structures. The original algorithm runs with two input parameters, $\epsilon$ and $MinPts$, defining ``neighborhood'' and ``type of points'', respectively. When using the algorithm to identify molecular clouds, we should adopt another threshold brightness temperature, $T_{mb, th}$, below which data is excluded. \cite{Yan2020} tested the effectiveness and robustness of the DBSCAN algorithm against different sets of $MinPts$ and $\epsilon$ using the MWISP data and recommended the most suitable parameters as $\epsilon=1$ and $MinPts=4$. The threshold $T_{mb, th}$ is set to 2$\sigma_{RMS}$. 

The spatial sampling of the velocity field, i.e., the number of projected pixels of the molecular cloud, is crucial for computing VSF (see discussions in section \ref{sec4.1}). Therefore, we selected 167 out of 2,214 molecular clouds with projected angular radii greater than 15 pixels in the \COl data for further VSF analysis. In the selected sample, the molecular clouds in the first quadrant lie in regions where $|b|\gtrsim 2$\arcdeg and $-5\leqslant v\leqslant 30$ km s$^{-1}$, which helps filter out distant clouds that are more concentrated in low-latitude regions \citep{Su2019} and allows us to assign the a ``near'' kinematic distance to each molecular cloud in the first quadrant.
For further analysis, the column density maps and some basic physical parameters such as mass, $M$, effective radius, $R_{\rm eff}$, and virial parameter, $\alpha_{\rm vir}$, of the selected molecular clouds are calculated using the same method as in \cite{Li2018} and \cite{Ma2021, Ma2022}. For the calculation, we adopt the kinematic distances of the molecular clouds derived using the Python routine ``Kinematic Distance Utilities''\footnote{\href{https://github.com/tvwenger/kd}{https://github.com/tvwenger/kd}}\citep{Wenger2018}, which uses the updated rotation curve of the Galaxy and Galactic rotation constants from model A5 in \cite{Reid2019}. 
As supplementary, the names of the molecular clouds and the derived physical parameters are presented in Table \ref{tabA1} in the Appendix. 

\begin{figure*}[htb!]
    \centering
    \includegraphics[trim=3.6cm 2.5cm 3.6cm 2.5cm, width= \linewidth, clip]{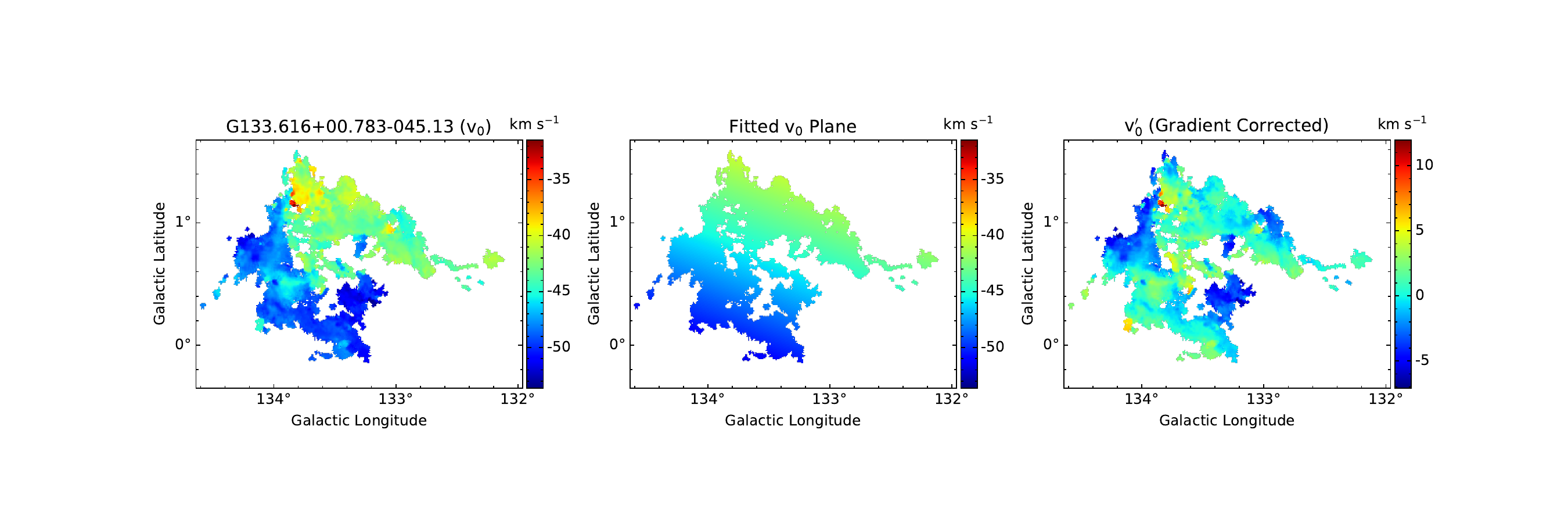}   
    \caption{Left: Intensity-weighted centroid velocity map of the $^{13}$CO structures identified within the boundary of the G133.616+00.783-045.13 molecular cloud. Middle: Fitted velocity plane of the centroid velocity map, representing large-scale global velocity gradients fitted through Eq. \ref{eq6}. Right: Gradient corrected velocity map used for VSF calculation.}
    \label{fig1}
\end{figure*}

\subsection{Turbulent Velocity Maps}\label{sec2.3}
The instantaneous velocity at a position in a fully developed turbulent field is the combination of a mean flow velocity, $\bar{v}$, and a fluctuation velocity $v'$ \citep{Frisch1995}. In observations, the profile of an optically thin molecular emission line can be considered a probability distribution of the radial velocities of all molecules along the given line of sight. The intensity-weighted centroid velocity (CV) of a spectrum, defined as the following formula, is the expectation of the radial velocity of the molecules along a given line of sight
\begin{equation}
\label{eq5}
    v_{\mathbf{x}} =  \bar{v}_{\mathbf{x}} + v'_{\mathbf{x}} = \frac{\int{T_i \cdot v_i dv}}{\int{T_i}dv},
\end{equation}
where $\mathbf{x}=(x, y)$ is the position vector, $i$ is the number of velocity channels, $T_i$ is the brightness temperature of the $i_{\rm{th}}$ channel, and $dv$ is the width of the velocity channel. The SF should be calculated from the $v'_{\mathbf{x}}$ field, meaning that the velocity should not contain systematic large-scale global motion. The velocity field of an ideally isolated turbulent molecular cloud is unaffected by any large-scale physical processes, therefore, $\bar{v}_{\mathbf{x}}$ should be uniform everywhere within the molecular cloud. However, molecular clouds in the Milky Way may possess global large-scale velocity gradients under the influence of various physical processes, such as circular motion around the Galactic center, stellar feedback, and gravitational collapse. To account for possible large-scale gradient which is not part of the internal turbulence, we fit each CV map with a two-dimensional first-order function with the form 
\begin{equation}
    \label{eq6}
    \bar{v}_{\rm fit} (x, y) = ax + by + c, 
\end{equation}
where $a$, $b$, and $c$ are fitting parameters. The fittings are implemented using the $scipy.optimize.curve\_fit$ routine, which is based on the non-linear least squares method. The parameters $a$ and $b$ could be approximately zero when no global gradient exists in the CV map. Then, the fluctuation velocity field is
\begin{equation}
    \label{eq7}
    v' = \bar{v} -\bar{v}_{\rm fit}.
\end{equation}
Although some previous observational studies calculated VSFs directly using the CV maps, we suggest that the gradient-corrected CV maps would be more reasonable. \cite{Stewart2022} also indicates that the gradient correction method provides better recovery of turbulent velocity dispersion.

Figure \ref{fig1} presents an example of the CV, the fitted velocity plane, and the gradient-corrected fluctuating velocity maps of the cloud G133.616+00.783-045.13, calculated according to the above methods. Although high-order gradients may still exist in the residual velocity map, we do not deal with them to avoid introducing additional artificial mistakes. 

\subsection{Calculation and Fitting of VSFs} \label{sec2.4}

High-order SFs are significantly influenced by the sampling completeness of the data and the intrinsic degree of intermittency of the turbulence \citep{Wit2004, DeMarco2017}. Therefore, in this work, we only calculate the first to third-order VSFs. The VSF is calculated according to Eq. \ref{eq1} with velocities derived through Eq. \ref{eq7}. The quantity $\mathbf{v'}_{\mathbf{x}} - \mathbf{v'}_{\mathbf{x+l}}$ is called the velocity increments at the spatial lag $\mathbf{l}$. We developed a Python program to compute the VSFs and only consider the azimuthal averaged VSFs across all directions. The algorithm input is a two-dimensional image, from which the algorithm selects all possible data pairs and records the corresponding increments and lags associated with these pairs. The increments and lags are then binned with the interval of one pixel. The $S_p(l)$ is the $p$-th order moment of the absolute velocity increments within the bin at $l$. The calculation of the SFs starts from $l = 2$ pixels, which approximately corresponds to the spatial resolution of the data. The largest spatial lag for VSF calculation is the distance between the farthest two pixels in the image. However, at any lags greater than $R_{\rm eff}$, the sampling for the velocity increments could be insufficient. Therefore, although we calculated the $S_p (l)$ for all possible spatial lags, discussions on the VSF shapes and power-law intervals are limited within $2\ \rm pixels <\it l\le R_{\rm eff}$.

In addition to the original VSFs, we also compute the column density-weighted VSFs as follows, similar to the definition of the density-weighted VSF in \cite{Chira2019}, 
\begin{equation}
    \label{eq8}
    S_{p, wei} (\mathbf{l}) = \frac{\langle N_{H_2(\mathbf{x})}N_{H_2(\mathbf{x+l})}|\mathbf{v'}_{\mathbf{x}} - \mathbf{v'}_{\mathbf{x+l}}|^p \rangle}{\langle N_{H_2(\mathbf{x})}N_{H_2(\mathbf{x+l})}\rangle},
\end{equation}
where ``$\langle \rangle$'' denotes average for all directions. 

An objective of this work is to test whether the velocity fields of Galactic molecular clouds are self-similar across scales, i.e., whether their VSFs present power-law shapes in some intervals of spatial lags, which should be the case according to the prediction of turbulence theories. Therefore, we fitted the VSFs of the 167 molecular clouds to examine the existence of power-law ranges. We use an ergodic and screening method to identify a common power-law range for $S_{p (p=1, 2, 3)}(l)$ and fit the VSF exponents $\zeta_{1, 2, 3}$ within this range for each molecular cloud. Specifically, we select all possible ranges for $l$ in each $S_p(l)-l$ curve within  $l\le R_{\rm eff}$ and perform a linear fit for $\log S_p$ and $\log l$. We record fitting ranges where the determination coefficient ($R^2$, the square of the Pearson correlation coefficient) is higher than 0.99 and choose the broadest recorded range as the candidate fitting range for each VSF. The overlapping range of the candidate fitting ranges for the 1st to 3rd-order VSFs is adopted as the common power-law range for the corresponding molecular cloud.

\section{Results} \label{sec3}
\subsection{Classification of VSFs According to the Significance of Power-law}\label{sec3.1}
\begin{figure*}[htb!]
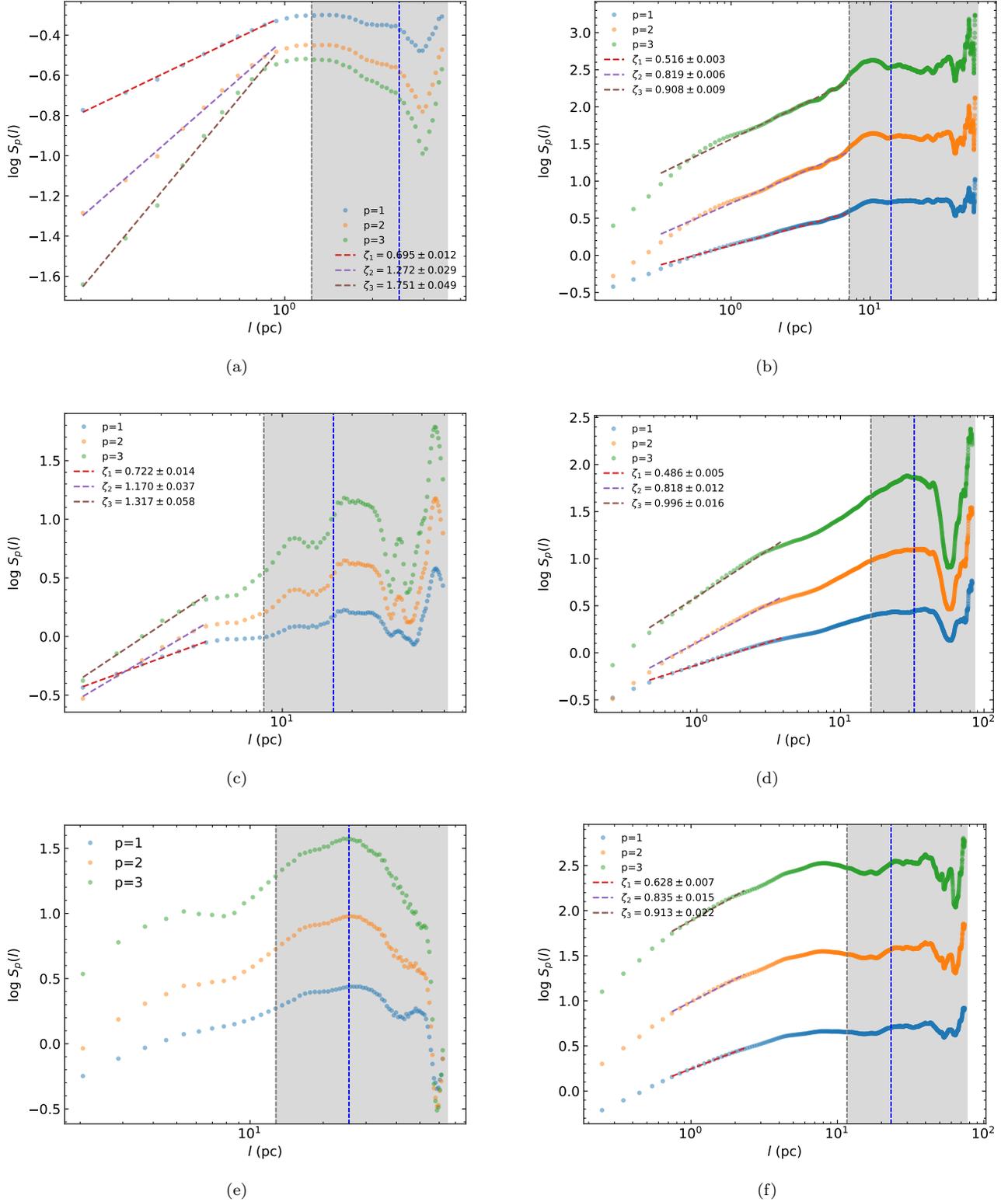

    \gridline{\fig{{G197.653-03.026+003.82_fit_sfs}.pdf}{0.45\textwidth}{(a)}
    \fig{{G141.256-02.038-003.51_fit_sfs}.pdf}{0.45\textwidth}{(b)}}
    \gridline{\fig{{G144.432+00.453-039.94_fit_sfs}.pdf}{0.45\textwidth}{(c)}
    \fig{{G202.028+01.592+005.99_fit_sfs}.pdf}{0.45\textwidth}{(d)}}
    \gridline{\fig{{G118.684+02.959-066.59_fit_sfs}.pdf}{0.45\textwidth}{(e)}
    \fig{{G134.761-00.382-007.65_fit_sfs}.pdf}{0.45\textwidth}{(f)}}
    \caption{Examples of VSFs in the ``S'' (a, b), ``M'' (c, d), and ``N'' (e, f) categories, respectively. The dashed lines overlaid on the VSFs in each panel represent the power-law fittings of the VSFs. The range of the fitted lines covers the common fitting range of the 1st-3rd order VSFs. The orders and the fitted power-law exponents of the VSFs are labeled in each panel. The vertical grey dashed line and blue dashed line show the lags at one and two times the effective radius, $R_{\rm eff}$, of the molecular cloud, respectively. The grey-shaded area highlights where lag$>R_{\rm eff}$, which is given for reference only.}
    \label{fig2}
\end{figure*}

Technically, we identified possible power-law ranges for all except one of the unweighted VSFs using the method in Section \ref{sec2.4}, although some possible power-law ranges only span two pixels. Indeed, VSFs derived from observations may not exhibit perfect power-laws across all scales in molecular clouds. Therefore, it is necessary to examine the significance of the VSFs being described by power laws. The molecular clouds in this study have different masses, sizes, and number of velocity sampling points. The absolute length of a power-law range is influenced by the size of the corresponding molecular cloud, making it not suitable to be a quantity to determine how significantly a VSF resembles a power-law. In the log-log $S_p(l)-l$ space, a sufficiently long segment that appears as a straight line should be considered indicative of a significant power law. Therefore, we define the significance of a power law in a VSF using the following criteria. 

If the endpoints of a power-law range are $l_1$ and $l_2$, and $(\log l_2 - \log l_1)/(\log R_{\rm eff} - \log \text{resolution}) \ge 0.7$, we define the VSF as significantly exhibiting a power-law distribution, denoted as a ``Significant (S)'' power-law. When $(\log l_2 - \log l_1)/(\log R_{\rm eff} - \log \text{resolution})$ is in the range [0.5, 0.7), the VSF is classified as a ``Moderate (M)'' power-law. If $(\log l_2 - \log l_1)/(\log R_{\rm eff} - \log \text{resolution}) < 0.5$, indicating that the $S_p(l)$ curve deviates from a power-law distribution in more than half of the range where $l \le R_{\rm eff}$, we consider that the VSF cannot be described by a power-law distribution and mark it as a ``None (N)'' power-law.

We classified the 167 molecular clouds into the S, M, and N categories according to the above criteria, with 100 clouds (60\%) in the S category, 40 clouds (24\%) in the M category, and 27 clouds (16\%) in the N category. The categories of each clouds are included in Table \ref{tabA1} in the Appendix. 
Two examples for each category are shown in Figure \ref{fig2}, with the first, second, and third rows corresponding to the S, M, and N categories, respectively. The left panels in each row show the VSFs of the molecular clouds with the smallest angular sizes in each category, while the right panels show those with the largest angular sizes. In Figure \ref{fig2}(a), the measured common power-law range for the three orders of VSFs is $\sim$0.2-0.8 pc, which is not as broad as the fitting range 0.5-4 pc of the VSFs in Figure \ref{fig2}(d). However, the physical radius and the spatial pixels of the molecular cloud corresponding to panel (a) are much smaller than those corresponding to panel (d). The measured power laws in the range of 0.2-0.8 pc in panel (a) provide good interpretations of the VSFs in panel (a), whereas the power laws in panel (d) are probably not a good description of its VSFs. The VSFs in panels (e)-(f) are considered to be none power-laws, and clearly, these VSFs do not resemble straight lines. In the shaded area in Figure \ref{fig2}, where $l>R_{\rm eff}$, we find that the VSFs exhibit considerable oscillations. These oscillations likely do not represent any physical processes but rather result from insufficient sampling of the velocity increments at large spatial lags.  

\begin{figure*}[htb!]
    \gridline{\fig{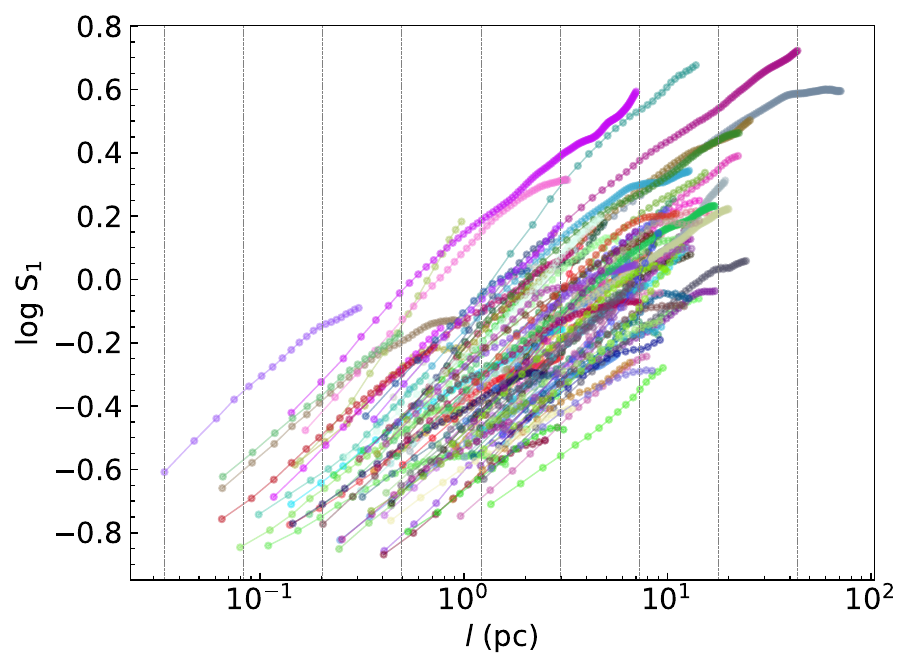}{0.33\textwidth}{(a)}
    \fig{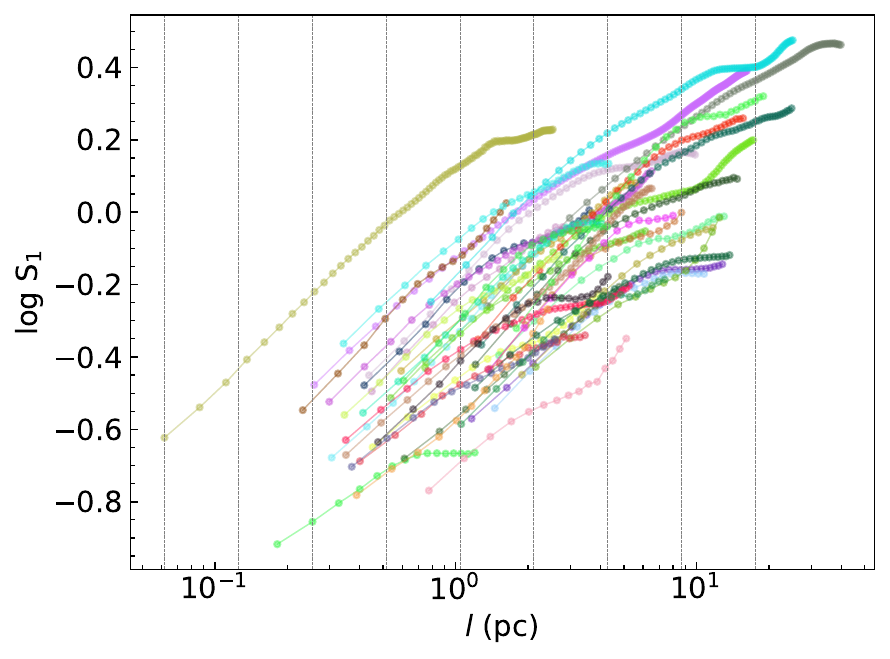}{0.33\textwidth}{(b)}
    \fig{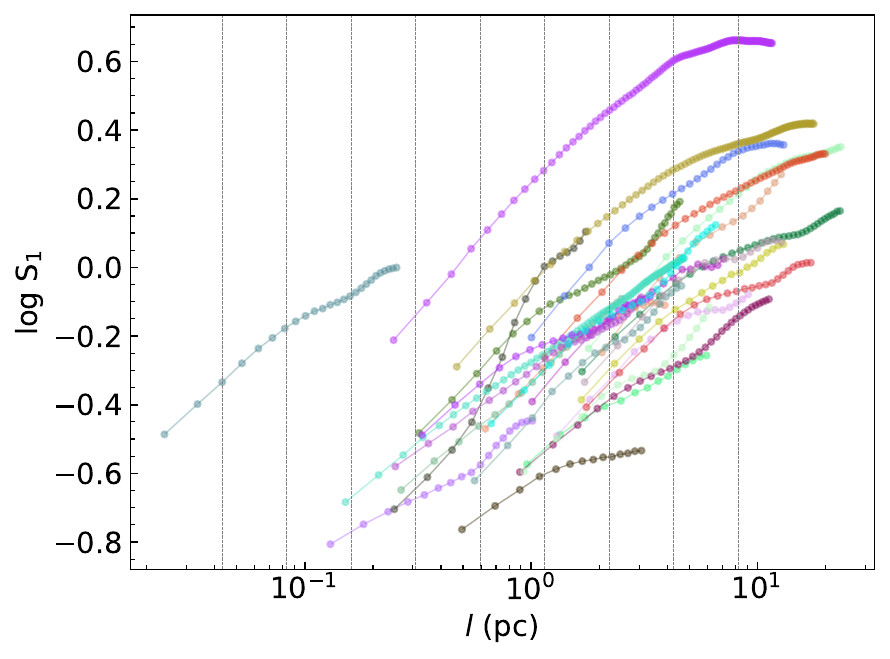}{0.33\textwidth}{(c)}}
    \gridline{\fig{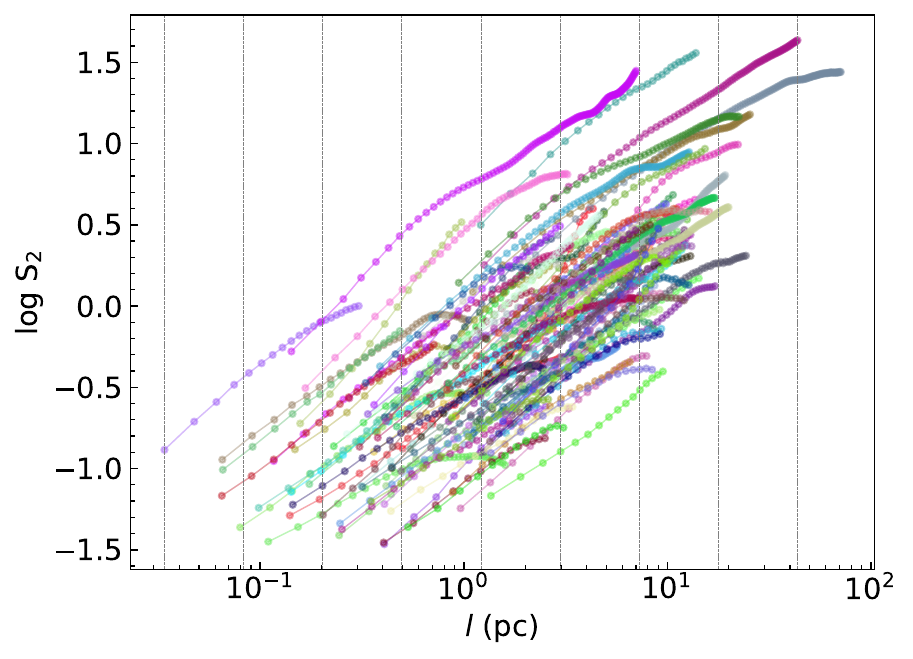}{0.33\textwidth}{(d)}
    \fig{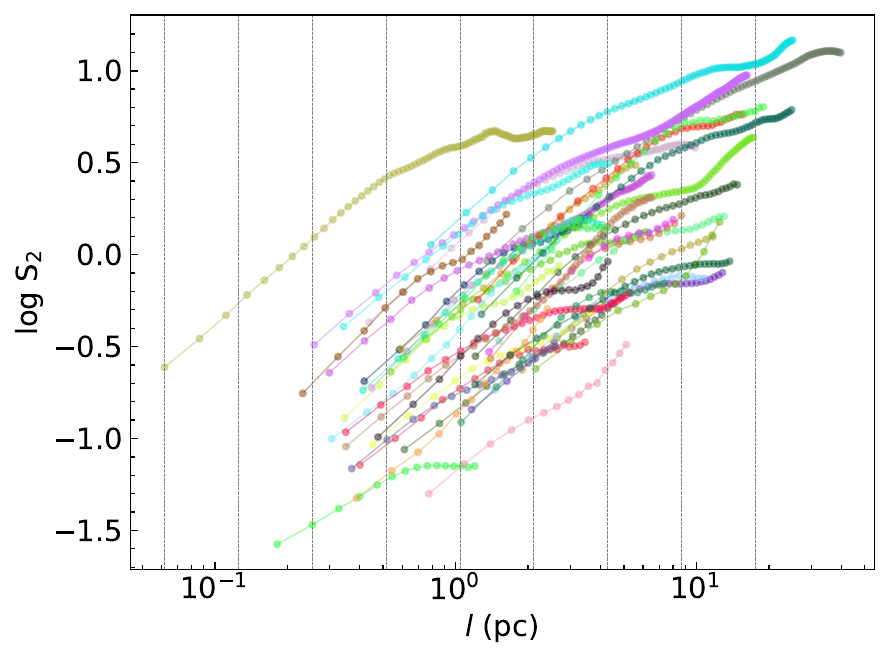}{0.33\textwidth}{(e)}
    \fig{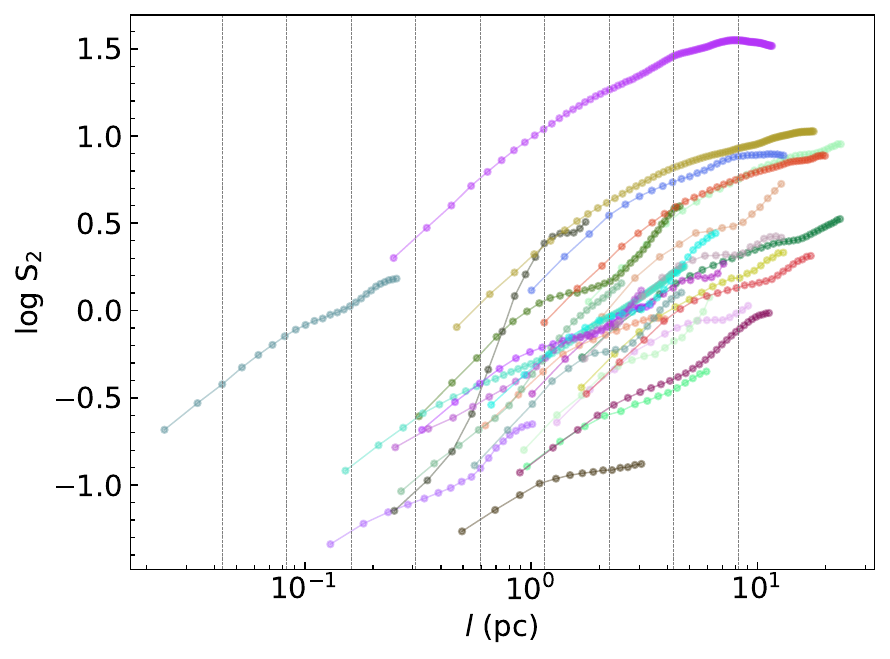}{0.33\textwidth}{(f)}}
    \gridline{\fig{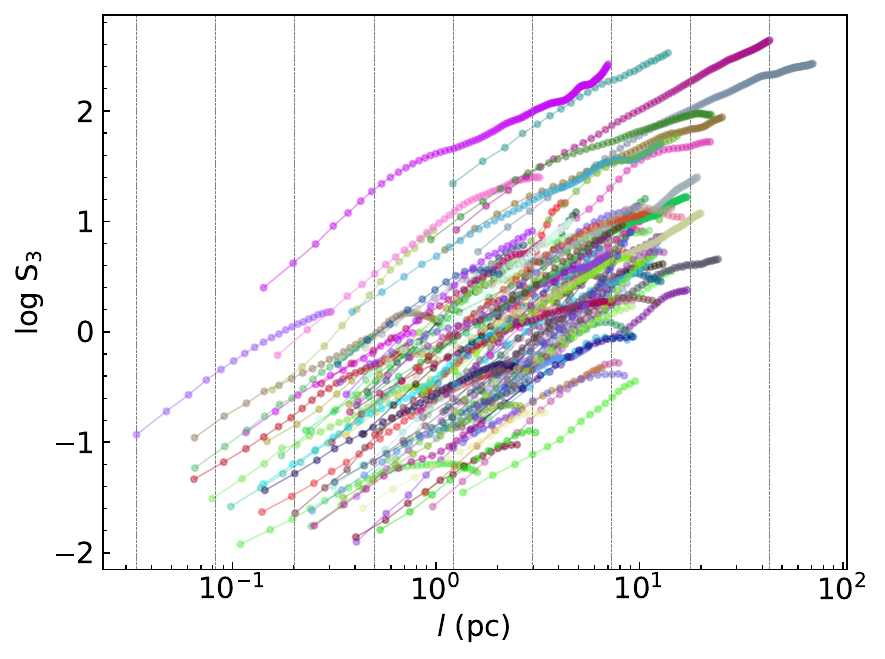}{0.33\textwidth}{(g)}
    \fig{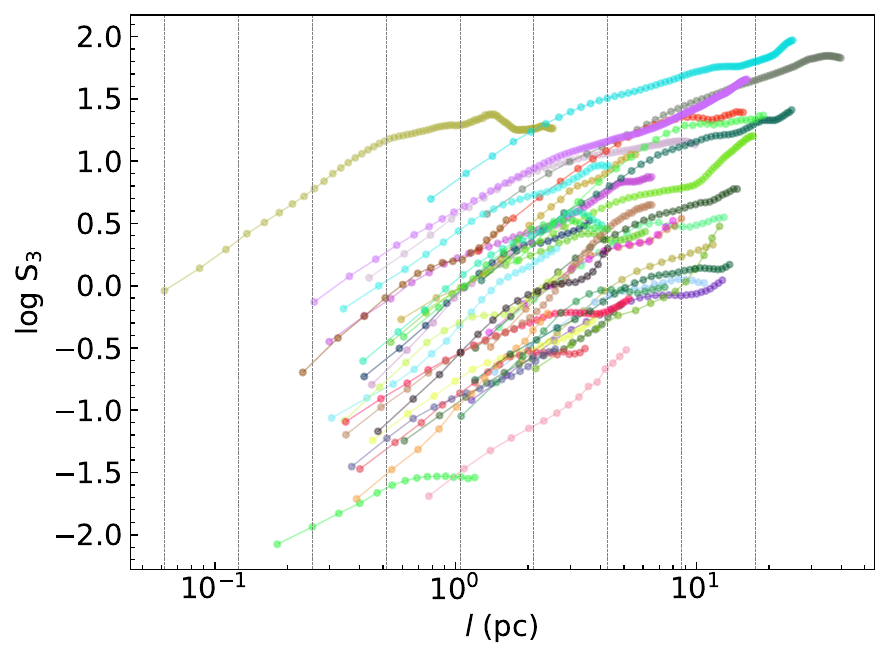}{0.33\textwidth}{(h)}
    \fig{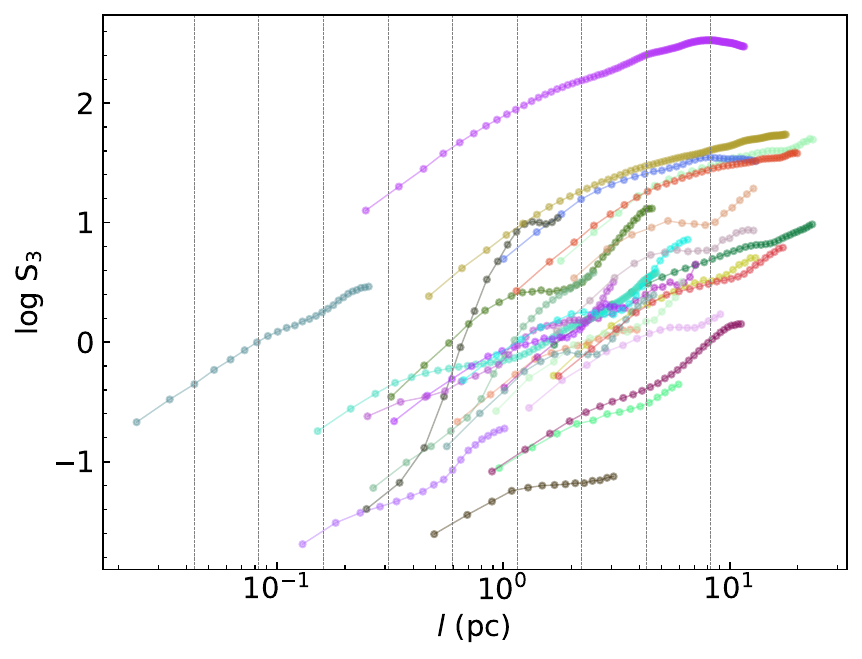}{0.33\textwidth}{(k)}}
    \caption{VSFs of the 167 molecular clouds, with the first to third rows corresponding to the first to third orders, respectively. From left to right, each column represents the S, M, and N categories, respectively. The colors of the VSFs are random, representing different molecular clouds, but within the first to third-order panels in each column, the VSF of the same molecular cloud is represented by the same color. The VSF for each molecular cloud is shown up to its $R_{\rm eff}$. The physical range that covers all power-law fitting intervals of the VSFs in each category is divided equally into eight bins on a logarithmic scale for future statistics.}
    \label{fig3}
\end{figure*}

Figure \ref{fig3} presents all the unweighted VSFs for the S, M, and N categories. Generally, all VSFs show an increasing trend with scale. The VSFs in the S category appear the most linear across all three orders, followed by the M category, while the least linear VSFs are observed in the N category. Although the VSFs in the N category still increase as a function of lag, they exhibit some fluctuations superimposed on the overall upward trend. Each category includes both large ($R_{\rm eff}>$10 pc) and small ($R_{\rm eff}<$1 pc) molecular clouds, indicating that whether the VSF exhibits a power-law distribution is independent of the size of the molecular cloud.

\subsection{Statistics of the Unweighted VSFs}\label{sec3.2}
\begin{figure}[htb!]
    \centering
    \includegraphics[width= \linewidth]{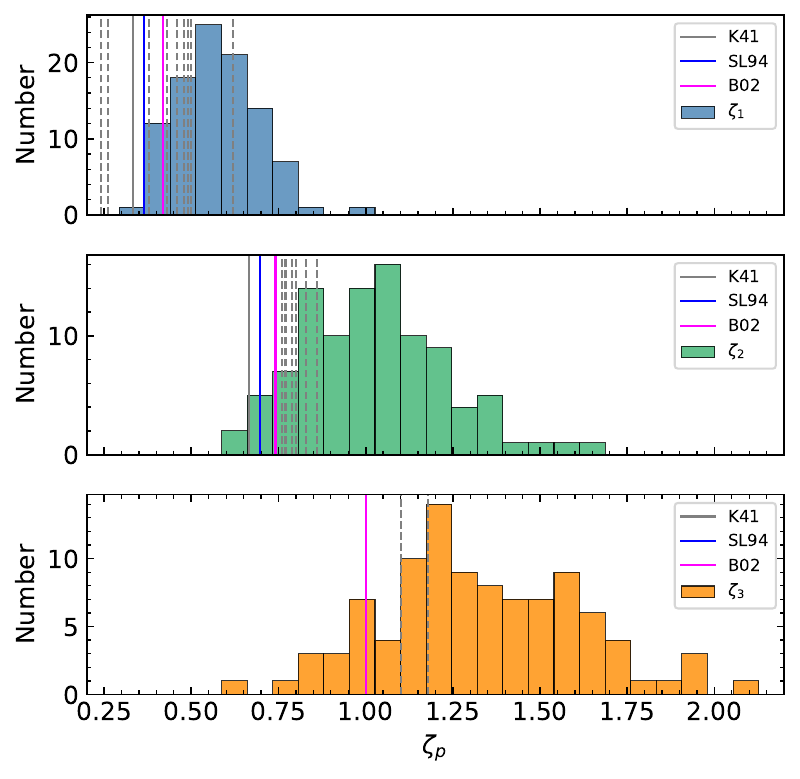}
    \caption{Histograms of the fitted power-law exponents of the VSFs in the ``S'' category, with the top to bottom panels corresponding to the first to third order VSFs, respectively. The theoretical predictions on the exponents are marked as vertical solid lines. The vertical dashed lines stand for the exponents obtained from historical observations as compiled by \cite{Chira2019}.}
    \label{fig4}
\end{figure}

Histograms of the fitted power-law exponents of VSFs in the S category are presented in Figure \ref{fig4} and summarized in Table \ref{tab1}. The median values of $\zeta_1$, $\zeta_2$, and $\zeta_3$ are 0.57, 1.01, and 1.30, with corresponding standard deviations of 0.11, 0.21, and 0.30, respectively. The measured range of power-law exponents covers the exponents obtained from previous observational studies as summarized in \cite{Chira2019}. Although, the exponents in this study are systematically slightly higher than earlier observational results for all orders of the VSF. Overall, the exponents significantly deviate from the theoretical predictions as we can see in Figure \ref{fig4} and Table \ref{tab1}. Comparatively, the absolute power-law exponents $\zeta_p$ are closest to the predicted values of \citetalias{Boldyrev2002a}, with $\zeta_1$, $\zeta_2$, and $\zeta_3$ differing from predictions by 38$\%$, 39$\%$, and 37$\%$, respectively. However, when considering the relative power-law exponent $Z = \zeta_p/\zeta_3$, the deviations from theoretical predictions become much smaller, with $Z_1$ and $Z_2$ differing from \citetalias{Boldyrev2002a} by 12$\%$ and 7$\%$, respectively.

\begin{deluxetable*}{cccccccccccc}[htb!]
    \tablewidth{700pt}
    % \tabletypesize{\tiny}
    \setlength{\tabcolsep}{12pt}
    \tablecaption{Summary of the Power-law Exponents of the unweighted and weighted VSFs in S Category\label{tab1}}
    \tablecolumns{4}
    \tablehead{
        \multirow{2}{*}{Orders} & \multicolumn{2}{c}{$\zeta_p$} & \multicolumn{2}{c}{$Z_p = \zeta_p/\zeta_3$} & \multicolumn{2}{c}{$\zeta_p$ (weighted)} & \multicolumn{2}{c}{$Z_p = \zeta_p/\zeta_3$ (weighted)} & \multicolumn{3}{c}{$\zeta_p$ (Model Predictions)} \\
         & 
        Mean & Std & Mean & Std & Mean & Std & Mean & Std  & K41 & SL94 & B02 
        }
    \startdata
    1 & 0.57 & 0.11 & 0.43 & 0.07 &  0.64 & 0.13 & 0.39 & 0.05 & 0.33 & 0.36 & 0.42 \\
    2 & 1.02 & 0.21 & 0.77 & 0.06 &  1.20 & 0.24 &  0.73 & 0.05 &  0.67 & 0.70 & 0.74 \\
    3 & 1.34 & 0.30 & -- & -- &   1.66 & 0.35 & -- & -- & 1.00 & 1.00 & 1.00 \\
    \enddata
    % \tablecomments{}
\end{deluxetable*}

\begin{figure*}[htb!]
    \gridline{\fig{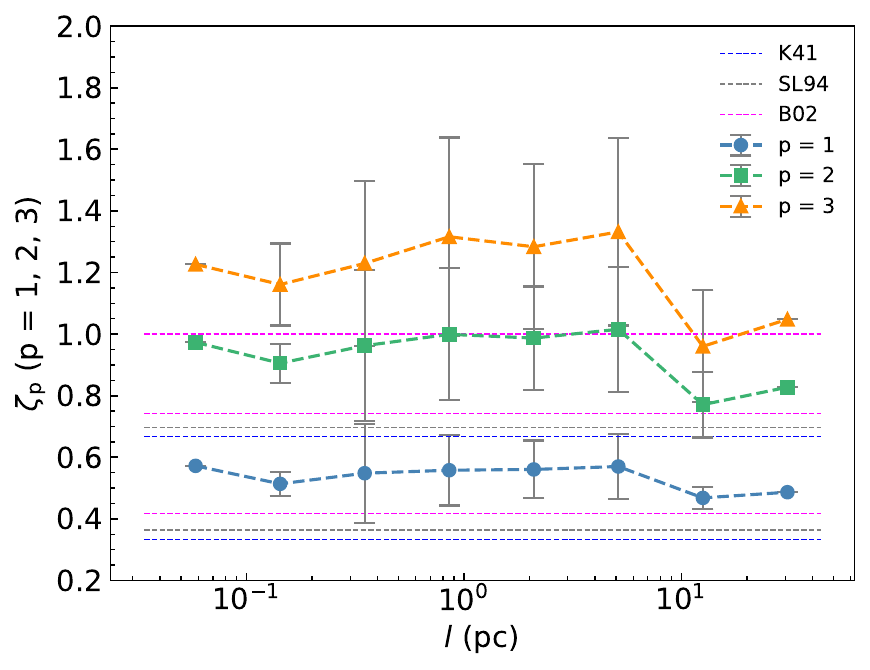}{0.49\textwidth}{(a)}
    \fig{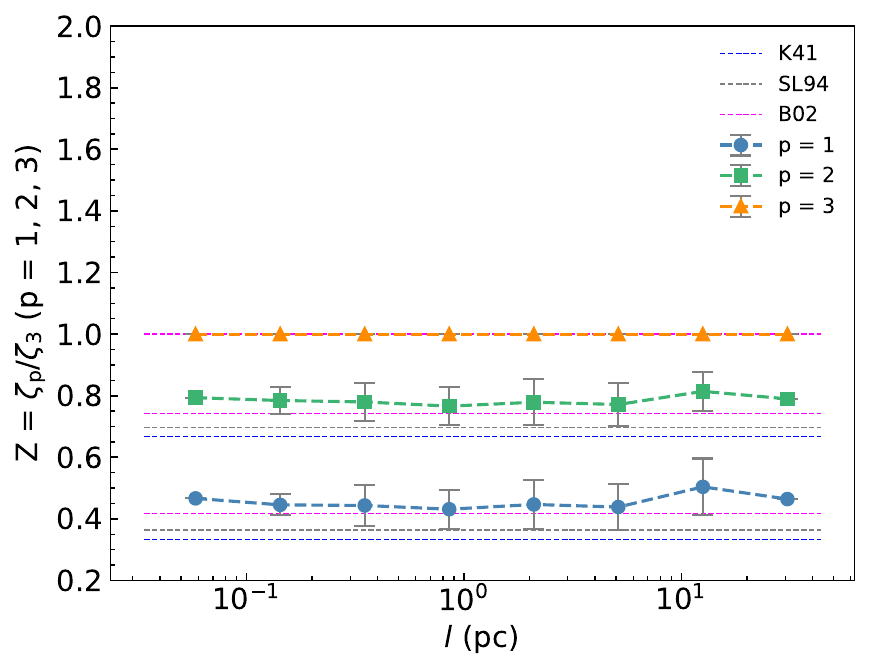}{0.49\textwidth}{(b)}}
    \caption{Variation of (a) $\zeta_p$ and (b) $Z_p$ as a function of spatial lags for the VSFs in the S category. The markers show the average $\zeta_p$ of the power-law exponents of the VSFs within the lag interval shown in Figure \ref{fig3}, with the corresponding error bars representing the standard deviations of the exponents. The horizontal dotted lines in panels (a) and (b) correspond to the theoretical predicted $\zeta_p$ and $Z_p$, respectively.}
    \label{fig5}
\end{figure*}

To account for the variations of the power-law exponents of the VSFs on different spatial scales, we divide the full range of the available scales into eight evenly spaced logarithmic intervals. We then calculate the average and standard deviation of the power exponents of the VSFs that follow power-law distributions within each interval. The intervals are marked by vertical dashed lines in Figure \ref{fig3}. The variations of $\zeta_p$ ($p = 1, 2, 3$) and $Z_p$ as a function of the spatial lags are shown in Figure \ref{fig5}. Only one molecular cloud exhibits a power-law VSF on scales smaller than 0.1 pc or greater than 20 pc. In contrast, all other spatial intervals contain at least two molecular clouds. Therefore, the two points at the smallest and largest scales in Figure \ref{fig5} do not have associated error bars. In the scale range from 0.1 to 5 pc, $\zeta_p$ generally does not exhibit a clear trend with varying scales and is systematically larger than the exponents predicted by the turbulence models. Additionally, the dispersion of the exponents mostly remains within 20$\%$ of the mean. However, at scales greater than ten pc, $\zeta_p$ decreases and approaches the \citetalias{Boldyrev2002a}'s predictions, especially for $\zeta_3$. Compared to the absolute scaling exponents of the VSFs, the relative scaling exponents, $Z_p$, lying much closer to the theoretical predictions of the \citetalias{Boldyrev2002a} model, being systematically higher by less than $\sim4\%$ across all scales from $\sim$0.06 to $\sim$30 pc. Similar figures for the molecular clouds in the M category are shown in Figure \ref{groups_M} in the Appendix. In Figure \ref{groups_M}, the power-law exponents show similar variation trends as those of the S category. However, both the deviations of the absolute exponents and the relative exponents from the theoretical predictions are slightly higher than the exponents of the S category. The results, to some extent, suggest that the molecular clouds with VSFs exhibiting significant power-laws express turbulent characteristics. However, a more robust observational depiction of other turbulence parameters, such as the intermittency exponent, requires accurate estimation of high-order structural functions, at least up to the sixth order. 

\subsection{Column density weighted VSFs}\label{sec3.3}
Considering the inhomogeneity of the column density in molecular clouds, we calculated the column density-weighted VSFs for the 167 molecular clouds according to Eq. \ref{eq8}. The significance of power-laws for the weighted VSFs is examined according to the criteria given in Section \ref{sec3.1}. There are 60 (36\%), 51 (31\%), and 56 (33\%) weighted VSFs classified as S, M, and N categories, respectively. Among the 60 molecular clouds in the S category, 47 are also classified into the S category based on their unweighted VSFs. The weighted VSFs of different categories are shown in Figure \ref{vsfs_wei} in the Appendix. In Figure \ref{vsfs_wei}, the column density-weighted VSFs, compared to the unweighted VSFs, exhibit fewer power-law distributions and more fluctuating features on different scales. In other words, the VSFs in Figure \ref{vsfs_wei}, especially those in the M and N categories, appear more curved. The statistics of the fitted scaling exponents for the weighted VSFs in the S category are summarized in Table \ref{tab1}. The column density-weighted VSFs show steeper power-laws than the unweighed VSFs. At small $l$, the weighting emphasizes the velocity differences within small-scale regions surrounding local maxima of column density, such as clumps or cores. In these regions, turbulence is expected to be more dissipated than in the low-density regions, implying that column density weighting reduces the p-th order moments of the velocity increments at relatively small lags. Contrarily, on large scales, the velocity increments between high column density regions and other locations tend to be more random, so the weighting does not significantly reduce $S_p$. According to \cite{Henshaw2020}, the lags corresponding to local minima in VSFs represent periodic spatial separations between internal structures like cores or clumps. This partly explains why the density-weighted VSFs have more fluctuation characteristics. The reason is that the minima in the weighted VSFs may represent regions of local maximum density, where the velocities of the molecular clouds also show local extreme values. 

The variation of the power-law exponents of the weighted VSFs against spatial lags is presented in Figure \ref{groups_S_wei} in the Appendix. On all scales, the power-law exponents of the weighted VSFs are larger than those of the unweighted VSFs. Although the absolute power-law exponents are significantly larger than the theoretically predicted values, as shown in Figure \ref{groups_S_wei}(a), the relative scaling exponents closely match the theoretical predictions across all possible spatial scales, even better than in the unweighted situations. 

\subsection{Signatures of Small-scale Intermittency} \label{sec3.4}
The consistency between the observed and theoretically predicted relative scaling exponents of the weighted and unweighted VSFs suggests that molecular clouds exhibit signatures of intermittent turbulence with hierarchical symmetry, with vortex sheets being the most intermittent structures \citepalias{Boldyrev2002a}. Generally, the intrinsic hierarchical symmetry of a velocity field can be examined using high-order VSFs. For example, the parameters $\beta$ and $\gamma$ in Eq. \ref{eq2} and \ref{eq3}, which define the fundamental properties of the turbulence, can be determined through $\beta$-test and $\gamma$-test \citep{She2001}. Therefore, to confirm that the statistical properties of the molecular cloud velocity fields conform to the intermittent hierarchical turbulence model, VSFs of at least six orders are required. Due to the uncertainties in measuring higher-order SFs, the discussion of higher-order VSFs is beyond the scope of this paper. In addition to quantitatively describing intermittency using higher-order SFs, intermittency can also be characterized by the statistics of velocity increments. At small scales, intermittency manifests as a non-Gaussian, long-tailed distribution of velocity increments \citep{Lis1996}.

\begin{figure*}[htb!]
    \gridline{\fig{{G197.653-03.026+003.82_inc_3d_4paper}.pdf}{0.4\textwidth}{(a)}
    \fig{{G141.256-02.038-003.51_inc_3d_4paper}.pdf}{0.4\textwidth}{(b)}}
    \gridline{\fig{{G144.432+00.453-039.94_inc_3d_4paper}.pdf}{0.4\textwidth}{(c)}
    \fig{{G202.028+01.592+005.99_inc_3d_4paper}.pdf}{0.4\textwidth}{(d)}}
    \gridline{\fig{{G118.684+02.959-066.59_inc_3d_4paper}.pdf}{0.4\textwidth}{(e)}
    \fig{{G134.761-00.382-007.65_inc_3d_4paper}.pdf}{0.4\textwidth}{(f)}}
    \caption{PDFs of the velocity increments at different lags of example molecular clouds in the ``S'' (a, b), ``M'' (c, d), and ``N'' (e, f) categories, respectively. The example molecular clouds are the same as those in Figure \ref{fig2}. The lag spans from its minimum to the $R_{\rm eff}$ of the molecular cloud. The lag range is equally divided into 30 intervals, and a corresponding velocity increment PDF is constructed for each interval.}
    \label{fig6}
\end{figure*}

\begin{figure*}[thb!]
    \gridline{\fig{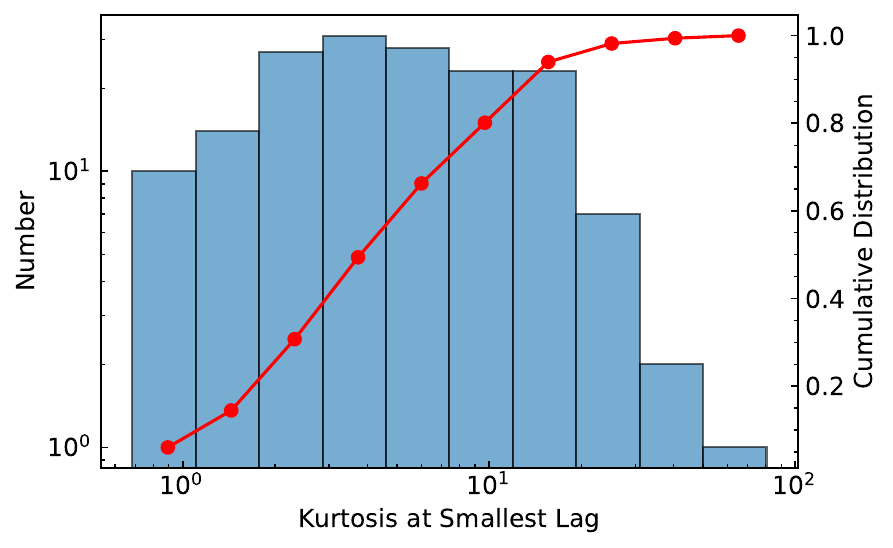}{0.495\textwidth}{(a)}
    \fig{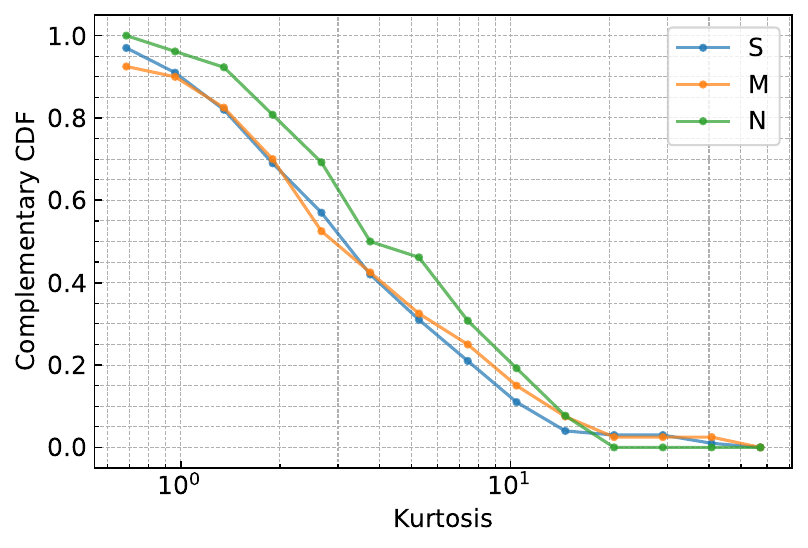}{0.45\textwidth}{(b)}}
    \caption{(a) Histogram of the kurtosis at the smallest lag of the 167 molecular clouds, overlaid with the cumulative distribution function of the distribution. (b) Complementary cumulative distribution function of the kurtosis for molecular clouds in the S, M, and N categories, respectively. Here, we employ the excess kurtosis.}
    \label{fig7}
\end{figure*}

The PDFs of the velocity increments for the VSFs in Figure \ref{fig3} are presented in Figure \ref{fig6}, showing two extreme cases, i.e., with a minimum and maximum number of sampling points, in each VSF category. Regardless of whether the VSFs exhibit significant power-law distributions, the velocity increment PDFs show deviations from Gaussian distributions on small scales, particularly at the smallest spatial lag. When the sample size is small (left column in Figure \ref{fig6}), the increment PDF at the smallest lag shows an exponential-like distribution. In contrast, with a large sample size (right column in Figure \ref{fig6}), the PDFs at the smallest lags exhibit pronounced features of sharp peaks and long tails. The ``tailedness'' of a distribution compared to a normal distribution can be quantified by excess kurtosis, $\kappa$, of the distribution. A value of $\kappa = 0$ indicates a normal distribution, positive $\kappa$ indicates overweighted tails and a sharp peak for the distribution, while negative $\kappa$ indicates thin tails and a flat peak. 

\begin{figure*}[htb!]
    \gridline{\fig{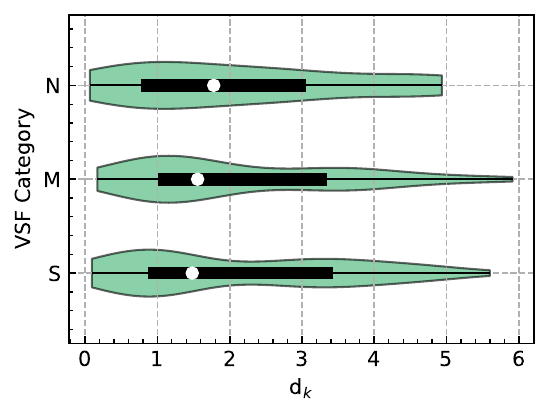}{0.33\textwidth}{(a)}
    \fig{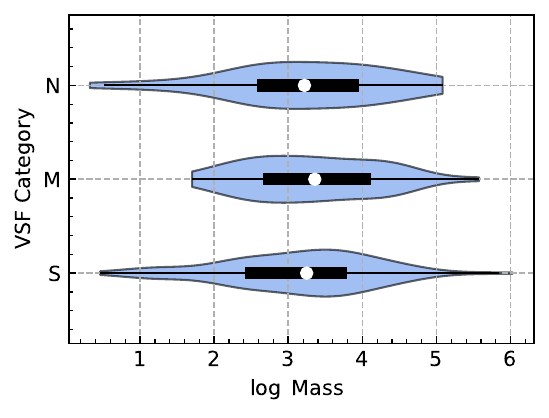}{0.33\textwidth}{(b)}
    \fig{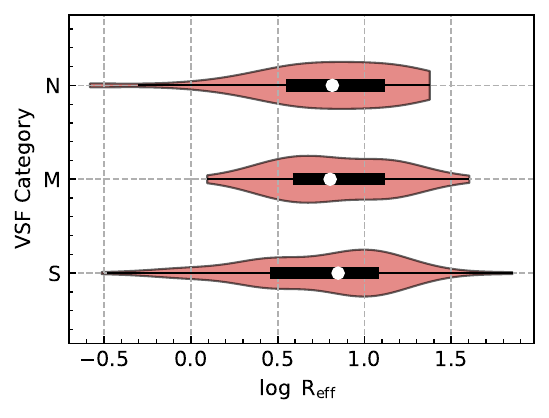}{0.33\textwidth}{(c)}
    }
    \caption{Violin plots of the (a) kinematic distance, (b) mass, (c) effective radius for molecular clouds in different VSF categories. The shaded areas represent the probability distribution of the corresponding physical quantity. White dots indicate the median values, thick black lines show the interquartile range (Q1 to Q3), and thin lines extend from Q1 - 1.5(Q3-Q1) to Q3 + 1.5(Q3-Q1), as long as these values lie within the range of the mass or radius.}
    \label{fig_violin}
\end{figure*}

We derive the mean of $\kappa (l = 2 \rm\ pixel)$ and $\kappa (l = 3\rm\ pixel)$ of the increments as the kurtosis at the smallest lag (hereafter, kurtosis) for each molecular cloud. The distribution of the kurtosis is shown in Figure \ref{fig7}. In our sample, the kurtosis of the velocity increments for all molecular clouds is greater than 0. Additionally, more than half of the molecular clouds have a kurtosis greater than $\sim$3, indicating that small-scale turbulence intermittency is widespread. Furthermore, as shown in Figure \ref{fig7}(b), it is evident that molecular clouds in the N category exhibit slightly stronger intermittency, i.e., higher kurtosis. The molecular clouds in the M category only show a slightly higher proportion at large kurtosis when compared to the S category. We also compared the distributions of kinematic distance, mass, and effective radius for molecular clouds in the three categories, as shown in the violin plots in Figure \ref{fig_violin}. In Figure \ref{fig_violin}(a), although the median $d_k$ of the three categories tends to increase sequentially, the differences among these values are negligible compared to the deviations within the distance distributions. Therefore, we can confirm that the classification of VSFs is unbiased by distance. The three categories also exhibit similar mass and radius distributions, particularly in terms of medians and quantiles. No significant correlation is found between non-power-law VSFs and the mass or radius of the molecular clouds. The stronger intermittency observed in molecular clouds with non-power-law VSFs cannot be straightforwardly attributed to any specific physical quantity.

%Therefore, the stronger intermittency for molecular clouds without apparent power-law VSFs cannot be straightforwardly attributed to a specific physical process. 

\section{Discussion} \label{sec4}
\subsection{Influence of Sampling on VSFs}\label{sec4.1}

Theoretically, the only inertial-range scaling relation at high Reynolds numbers that can be strictly derived from the Navier-Stokes equation is Kolmogorov's 4/5th law, resulting in $\zeta_3=$ 1 \citep{Frisch1995}. Therefore, if molecular clouds are intrinsically dominated by turbulence, a plausible expectation is that the observed power-law exponents of the third-order structure functions should not deviate much from unity. 

\begin{figure}[htb!]
    \centering
    \includegraphics[width = \linewidth]{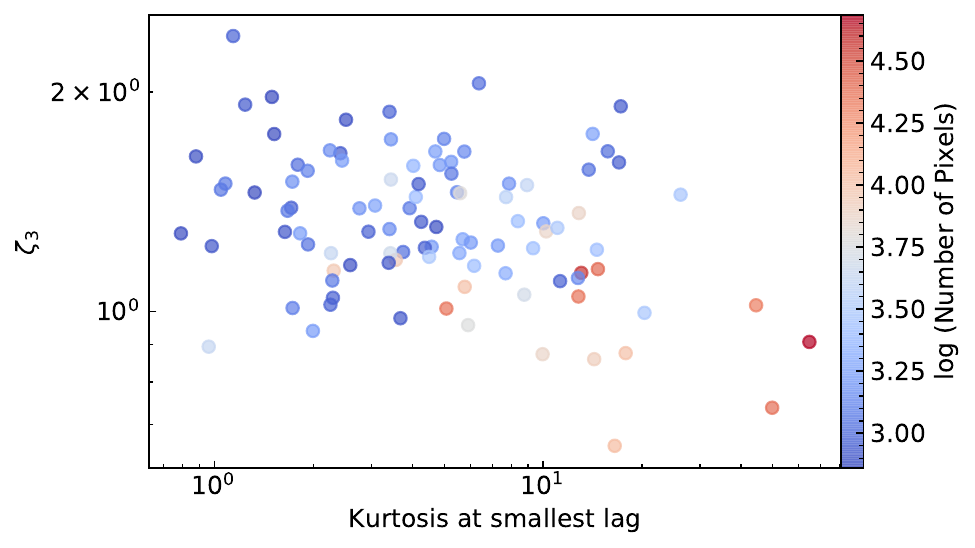}
    \caption{Relationship between the power-law exponents of the third-order VSFs and the kurtosis of velocity increments at smallest lags. The markers are weighted by the number of spatial pixels of the molecular clouds.}
    \label{fig9}
\end{figure}

We observed a relatively large dispersion in the power-law exponents of the third-order VSFs, more or less deviating from one. This deviation may be related to the intermittency of molecular clouds on small scales and the limited sampling of velocities in our observations. To verify this, we present the relationship between $\zeta_3$ and the kurtosis of velocity increments on small scales ($l$ = 2-3 pixels) for molecular clouds in the S category in Figure \ref{fig9}, with data points weighted by the number of projected pixels of the molecular clouds. Although with a large scatter, a trend of an anti-correlation between $\zeta_3$ and the kurtosis is clear in Figure \ref{fig9}. Another trend observed in Figure \ref{fig9} is that as the number of pixels in the molecular cloud increases, the kurtosis of the velocity increments at small lags becomes larger, and $\zeta_3$ becomes smaller, approaching one. 

As suggested by \cite{DeMarco2017}, when intermittency is present, the velocity increment PDFs exhibit strong non-Gaussian behavior at small lags, possibly following a symmetric normal inverse Gaussian distribution. The stronger the intermittency, the more pronounced the heavy tails of the velocity increment PDF, requiring a large number of samples to capture the low-probability events of large velocity increments. Without sufficient sampling, the calculated moments of the increments could be underestimated, especially for high-order (p$>$3) structure functions. The intrinsic kurtosis at small lags of each molecular cloud is unknown. At a fixed spatial resolution, we cannot assert whether the kurtosis of a specific molecular would increase or decrease if we sample its velocity field with higher spatial resolutions. However, from the statistical results in Figure \ref{fig9}, for those molecular clouds having relatively large $\zeta_3$ and small kurtosis at small lags, their pixel numbers are also relatively small, implying that the deviation of $\zeta_3$ from one is partly due to insufficient sampling. 

\subsection{Comparison with Numerical Simulations} \label{sec4.2}

\begin{figure}[thb!]
    \centering
    \includegraphics[width = \linewidth]{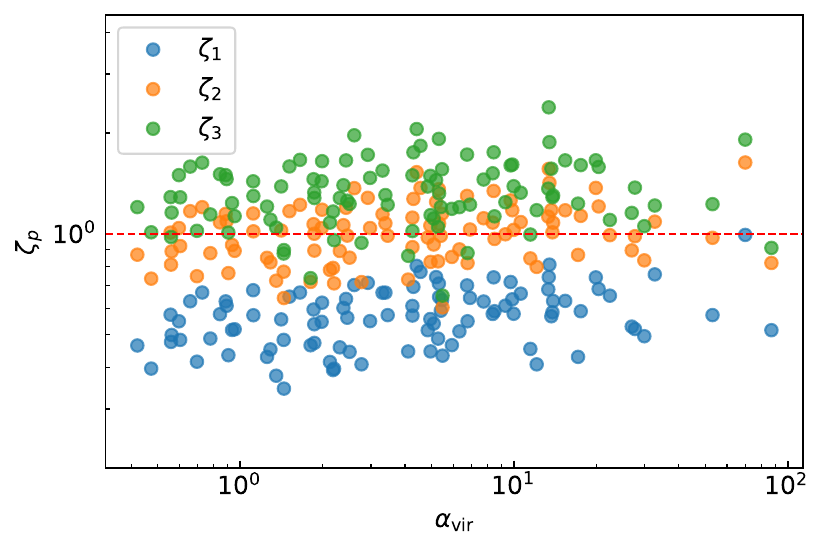}
    \caption{Relationship between the power-law exponents of the first- to third-order VSFs and the virial parameters of the molecular clouds in the S category.}
    \label{fig10}
\end{figure}

\cite{Chira2019} found in their simulations that the VSFs of molecular clouds are significantly influenced by gravitational contraction, which can drive small-scale motions and alter the scaling exponents of VSFs within the range of $\sim$ 0.8 to 8 pc, even changing them from positive to negative. However, we did not observe any power-law VSFs with negative exponents for either weighted or unweighted VSFs. The relationship between the measured scaling exponents of the VSFs and the virial parameters, which partly reflect the importance of gravity to molecular clouds, is given in Figure \ref{fig10}. No clear correlation between the exponents and the virial parameters was observed. 

\cite{Chira2019} also discussed why the observed VSFs always exhibit positive power-law exponents. They suggested that this is because the commonly used observational tracers, such as CO and its isotopologues, are affected by optical depth and thus cannot trace dense, gravitationally dominant, collapsing regions of molecular clouds. However, in \cite{Ma2021}, we found that the molecular clouds in the second quadrant, traced by $^{13}$CO, have number densities of $\sim$ 700-1000 cm$^{-3}$, which is comparable to the densities of the molecular clouds in \cite{Chira2019}. 
Additionally, although opacity broadening can affect line profiles, it does not significantly impact centroid velocity measurements since line saturation is symmetric. Therefore, we propose that gravity-dominated regions should remain confined to small scales within molecular clouds. Otherwise, unless the gravity-dominated gas reaches such high densities on scales of 0.8-8 pc that the $^{13}$CO emission is completely obscured (which is not observed in practice), we should still be able to detect a change in the power-law exponents of the VSFs in our observations. Our inference is consistent with the numerical simulations by \cite{Hu2022}, which indicate that the influence of gravity on the VSFs is mainly concentrated on clump scales, i.e., sub-pc scales. However, due to the complexity of VSF analysis, more detailed comparisons between observations and numerical simulations are still needed to confirm the gravity-dominated scales of molecular clouds. To directly verify from observations whether the velocities in gravity-dominated dense regions can significantly change the power-law exponents of the VSFs, we still need high-density tracers with high volume-filling factors in the observed area, which may necessitate very deep observations. 

\subsection{Implications on Driving Sources of Turbulence}\label{sec4.3}

As introduced in Section \ref{sec1}, \cite{Heyer2004} and \cite{Brunt2009} identified consistent power-law scaling relations for hundreds of Galactic molecular clouds, suggesting that large-scale external forces drive turbulence and that there is a uniform formation mechanism for these clouds. This is understandable, as one of the key features of turbulence is the loss of large-scale asymmetry and directional information after cascading into the inertial scale interval \citep{Frisch1995}. Therefore, the local or internal driving of turbulence should be somewhat reflected in the non-power-law behavior of the VSFs.

Based on our current results, at least 60\% of molecular clouds exhibit turbulent characteristics in their velocity fields, as indicated by the presence of a significant power-law interval in the VSF and the consistency between the measured power-law indices and the predictions of \citetalias{Boldyrev2002a} model. It should be noted that this percentage represents a lower limit of power-law VSFs since we define a substantial power-law with a stringent criterion (see Section \ref{sec3.1}). For different molecular clouds, the power-law scaling exponents of the 1st to 3rd-order VSFs exhibit deviations within 20\% of their mean values, consistent with the results from \cite{Heyer2004}. However, in contrast to both narrow ranges of the scaling exponent and coefficient reported by \cite{Heyer2004}, our work shows that the scaling coefficient, $\nu_0$, of the VSF has a relatively broad distribution, as reflected by the spread of the VSFs along the Y-axis in Figure \ref{fig3}. In this study, the mean value of $\nu_0$ is 0.47 with a standard deviation of 0.28, resulting in $\sigma_{\nu_0}/\langle \nu_0 \rangle = 59\%$, which is significantly higher than the deviation of $\nu_0$ ($\sim$10-20\%) reported in \cite{Heyer2004}. This discrepancy likely arises from the different algorithms used for VSF calculations in \cite{Heyer2004} and our study. \cite{Heyer2004} used the PCA and the autocorrelation function techniques to compute the VSFs, leading to normalized velocity dispersions and, consequently, a narrower range of the scaling coefficient of the VSFs. Therefore, similar to the findings of \cite{Heyer2004} and \cite{Brunt2009}, the consistency of power-law scaling exponents and the sufficiently long spanning range of power-law intervals for the molecular clouds in the S category in our study further support the picture of large-scale external driving of turbulence in these molecular clouds. However, the large dispersion in the scaling coefficient $\nu_0$ indicates that the turbulence energy should vary across different molecular clouds. In terms of energy availability, \cite{MacLow2004} proposed that background supernovae might be the primary drivers of interstellar turbulence, compared to other large-scale mechanisms such as magnetorotational instability, gravitational instability, and stellar winds. However, the energy input sources for different molecular clouds in our sample are indistinguishable from the current results.

\begin{figure}[htb!]
\gridline{\fig{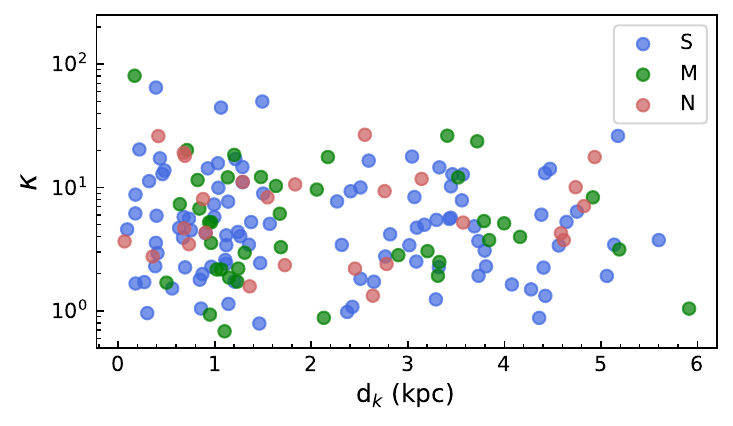}{0.45\textwidth}{(a)}}
    \gridline{\fig{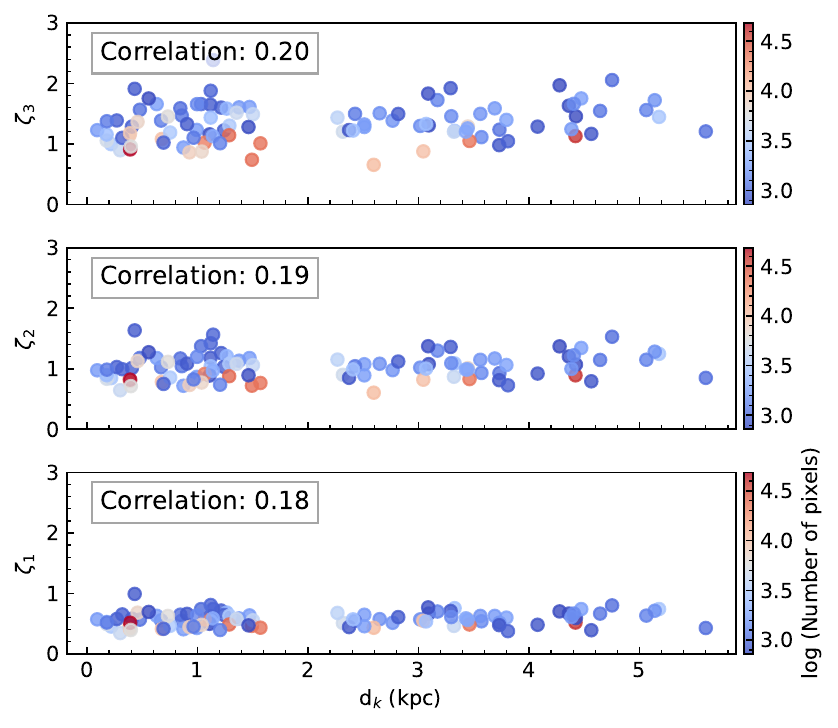}{0.49\textwidth}{(b)}}
    \caption{Relationship between (a) kurtosis ($\kappa$) and the kinematic distance for the three categories of molecular clouds and between (b) the power-law exponents of the first- to third-order VSFs and the kinematic distances for molecular clouds in the S category. The Pearson correlation coefficients for each dataset in subfigure (b) are indicated in respective panels.}
    \label{fig11}
\end{figure}

Another significant finding from our work, which differs from \cite{Heyer2004}, is that approximately 40\% of the VSFs exhibit deviations from power-law distributions. In Section \ref{sec3}, we also observed that molecular clouds in the N category display stronger intermittency at small spatial scales. In Figure \ref{fig_violin}, we have found that the existence of power-law distributions in VSFs is independent of the distances of molecular clouds. In addition to checking the heliocentric distances, we also carefully checked the Galactocentric distances of the molecular clouds and found that the distribution of molecular clouds in the Galactic plane for the three categories appears random, indicating that the classification of VSFs and the slightly stronger intermittency of the N-type molecular clouds is possibly independent of Galactic environments. The $\kappa$-$d_k$ relationship for the three categories of molecular clouds and the $\zeta_p$-$d_k$ relationship for the S category are presented in Figure \ref{fig11}. In this figure, no significant correlations are observed between the kurtosis and the distance or between the power-law exponent and the distance. Besides, molecular clouds with fewer pixel numbers tend to exhibit larger fitted slopes for VSFs, as discussed in Figure \ref{fig9}.

So far, the velocity intermittency likely shows no correlation with the mass, size, distance and Galactic environment of molecular clouds. However, the internal kinematics of molecular clouds are highly complex and involve various physical processes. Energy injection or the presence of structures with periodic or regular spatial distributions can significantly impact the power-law behavior of the VSF \citep{Frisch1995, Chira2019, Henshaw2020}. Notably, some individual molecular clouds in this study with VSFs that do not exhibit significant power-law behavior show active star-forming activities, such as the Rosette, Sh2-152, and W3-W4-W5 GMCs. Therefore, the relationship between turbulence intermittency and stellar feedback, i.e., local environment of molecular clouds, requires further investigation. This is especially relevant for M- and N-type molecular clouds, which we aim to explore in a future paper.

\section{Conclusion} \label{sec5}
In this work, we present a systematic analysis of the VSFs of 167 molecular clouds with angular sizes greater than $\sim$176 arcmin$^2$ in three sectors of the Galactic mid-plane. We calculated the 1st- to 3rd-order VSFs and examined whether the VSFs present power-law distributions since power-law VSFs are possible signatures of turbulence cascading in the molecular clouds. We also fit the power-law scaling exponents of the VSFs for molecular clouds whose 1st-3rd order VSFs all exhibit power law distributions and compare the fitted exponents with the predictions of turbulence theory. By constructing velocity-increment PDFs at different spatial lags, we quantitatively estimate the intensity of the small-scale intermittency of turbulence for the molecular clouds. The main findings are summarized as follows. 

1. About 60\% of the unweighted VSFs of the molecular clouds exhibit significant power-law distributions, while 24\% show moderate power-laws over relatively narrow spatial intervals of lags, and 16\% show no power-law intervals. The fitted relative scaling exponents, $Z=\zeta_p/\zeta_3$ (p = 1, 2), for the power-law VSFs, are generally consistent with the \citetalias{Boldyrev2002a} intermittency model for compressible turbulence in the ISM. Across different molecular clouds, the power-law intervals of unweighted VSFs cover a scale range from $\sim$0.1 to $\sim$10 pc, in which the dispersion of the absolute power-law scaling exponents is only 20\% of their mean, while the dispersion of the relative scaling exponents is only 4\%. 

2. When considering column density weighted VSFs, the proportion of power-law VSFs is reduced to 36\%. The power-law of VSFs becomes steeper after column density weighting, implying that the turbulent energy is more dissipative in small-scale high-density regions than in diffuse areas. The relative power exponents of the weighted VSFs almost perfectly recover the predicted values of the \citetalias{Boldyrev2002a} turbulence theory. Compared to the unweighted VSFs, the column density weighted VSFs present more fluctuations on relatively large spatial lags, which may reflect consistent periodic separations of velocity and column density local maxima or minima.  

3. The velocity fields of all molecular clouds show small-scale intermittency in terms of the velocity increment PDF, which is consistent with \citetalias{Boldyrev2002a} being a model that describes intermittent turbulence. Molecular clouds with unweighted VSFs that do not follow power-law distributions exhibit slightly stronger intermittency than those with moderate or significant power-law VSFs. However, regardless of whether the VSF exhibits a power-law distribution, other properties of molecular clouds in different categories, such as mass and size, do not show significant differences. 

4. In our sample, negative VSF exponents indicative of gravitational collapse, which would cause large relative infalling velocities on small scales, are not observed. Furthermore, the scaling exponents of the VSFs do not correlate with the gravitational states of the molecular clouds. These results suggest that gravity-dominated scales in molecular clouds still need further investigation.

5. The widespread power-law distribution and relatively consistent scaling exponents for molecular clouds in the S category suggest that turbulence within these molecular clouds is likely driven by external large-scale processes, while the scattered scaling coefficients may suggest different energy sources. About 40\% of VSFs show certain degrees of deviation from power law distributions, indicating that the influence of local environments on the internal turbulence of molecular clouds may not be negligible.

\begin{acknowledgments}
    We thank the anonymous referee for the constructive comments that help to improve this manuscript. This research made use of the data from the Milky Way Imaging Scroll Painting (MWISP) project, which is a multi-line survey in $^{12}$CO/$^{13}$CO/C$^{18}$O along the northern galactic plane with PMO-13.7 m telescope. We are grateful to all the members of the MWISP working group, particularly the staff members at PMO-13.7 m telescope, for their long-term support. MWISP was sponsored by National Key R$\&$D Program of China with grants 2023YFA1608000 $\&$ 2017YFA0402701 and by CAS Key Research Program of Frontier Sciences with grant QYZDJ-SSW-SLH047. H.W. acknowledges the support of NSFC grant 11973091. Y.M. acknowledges the support of NSFC grants 12303033 and 11973090. M. Z. acknowledges the support of NSFC grants 12473026 and 12073079.
\end{acknowledgments}
    
\appendix
\section{Supplementary Table}
Table \ref{tabA1} presents the physical parameters of the molecular clouds and also gives the left and right fitting range of the VSFs, the fitted power-law exponents, $\zeta_1$, $\zeta_2$, and $\zeta_3$ of the unweighted VSFs, kurtosis of velocity increments at the smallest lags and the categories of molecular clouds. 
\begin{longrotatetable}
	\begin{deluxetable*}{lrrrrrrrrrrrrrrc}
		\tablewidth{10pt}
		\tabletypesize{\tiny}
		 \tablecaption{Physical Paramteters of the Molecular Clouds and VSF Related Paramteters\label{tabA1}}
		 \tablecolumns{17}
		 \tablehead{
			 \colhead{Name} & \colhead{$l$}& \colhead{$b$}& \colhead{$v$} &\colhead{$d_{k}$} & \colhead{Mass} & \colhead{$R_{\rm eff}$} & \colhead{$\alpha_{\rm vir}$} & \colhead{N$\rm_{pix}$} & \colhead{$\kappa$} & \colhead{Left} & \colhead{Right} & \colhead{$\zeta_1$} & \colhead{$\zeta_2$} &\colhead{$\zeta_3$} & \colhead{Type}\\
			&\colhead{(\arcdeg)} & \colhead{(\arcdeg)}& \colhead{(km s$^{-1}$)} & \colhead{(kpc)} & \colhead{(M$_{\sun}$)}  & \colhead{(pc)} &  &  & &\colhead{(pc)} &\colhead{(pc)}& & &  & \\
			\colhead{(1)} & \colhead{(2)} & \colhead{(3)} & \colhead{(4)} & \colhead{(5)} & \colhead{(6)} & \colhead{(7)} & \colhead{(8)} & \colhead{(9)} & \colhead{(10)} & \colhead{(11)} & \colhead{(12)} & \colhead{(13)}& \colhead{(14)} & \colhead{(15)} & \colhead{(16)} 
			}
		 \startdata
		MWISP G202.028$+$01.592$+$005.99 & 202.028&1.592&5.99 & 0.712 & 37499.4 &  16.299 & 2.31 & 77828 & 20.14 & 0.466&3.883&0.486 & 0.818&0.996 & M\\
MWISP G141.256$-$02.038$-$003.51 & 141.256&-2.038&-3.51 & 0.391 & 1681.2 &  7.059 & 87.14 & 48408 & 64.60 & 0.313&6.796&0.516 & 0.819&0.908 & S\\
MWISP G134.761$-$00.382$-$007.65 & 134.761&-0.382&-7.65 & 0.679 & 5846.8 &  11.645 & 39.42 & 43678 & 19.12 & ...&...&... & ...&... & N\\
MWISP G110.586$+$00.170$-$050.27 & 110.586&0.170&-50.27 & 4.421 & 1063805.0 &  71.273 & 0.95 & 38598 & 13.07 & 1.608&26.685&0.519 & 0.890&1.129 & S\\
MWISP G145.817$+$03.428$-$000.42 & 145.817&3.428&-0.42 & 0.170 & 170.3 &  2.549 & 45.11 & 33382 & 80.53 & 0.062&0.705&0.638 & 1.077&1.252 & M\\
MWISP G207.156$-$01.984$+$012.95 & 207.156&-1.984&12.95 & 1.290 & 39030.7 &  17.909 & 2.88 & 28623 & 11.12 & ...&...&... & ...&... & N\\
MWISP G224.065$-$01.303$+$016.32 & 224.065&-1.303&16.32 & 1.287 & 48080.5 &  17.244 & 0.78 & 26660 & 14.68 & 0.468&17.128&0.487 & 0.876&1.143 & S\\
MWISP G122.500$-$00.695$-$017.92 & 122.500&-0.695&-17.92 & 1.493 & 29162.8 &  19.287 & 1.81 & 24784 & 49.77 & 0.760&13.789&0.465 & 0.719&0.738 & S\\
MWISP G118.998$+$03.033$-$018.01 & 118.998&3.033&-18.01 & 1.570 & 44885.8 &  20.008 & 0.90 & 24120 & 5.08 & 0.571&19.980&0.434 & 0.763&1.009 & S\\
MWISP G111.529$-$02.563$-$039.09 & 111.529&-2.563&-39.09 & 3.462 & 170624.9 &  43.897 & 5.28 & 23877 & 12.81 & 1.259&43.555&0.487 & 0.827&1.048 & S\\
MWISP G106.514$+$04.083$-$007.01 & 106.514&4.083&-7.01 & 1.065 & 16548.4 &  12.925 & 4.26 & 21875 & 44.47 & 0.387&7.513&0.570 & 0.913&1.019 & S\\
MWISP G216.584$-$02.515$+$025.52 & 216.584&-2.515&25.52 & 2.172 & 22913.9 &  25.142 & 10.21 & 19899 & 17.67 & 2.053&12.478&0.428 & 0.590&0.624 & M\\
MWISP G115.611$-$02.718$-$002.06 & 115.611&-2.718&-2.06 & 0.416 & 966.2 &  4.779 & 6.86 & 19597 & 26.12 & ...&...&... & ...&... & N\\
MWISP G133.616$+$00.783$-$045.13 & 133.616&0.783&-45.13 & 3.719 & 375295.8 &  40.165 & 0.61 & 17322 & 23.69 & 1.893&17.579&0.514 & 0.810&0.888 & M\\
MWISP G144.531$-$01.229$-$029.14 & 144.531&-1.229&-29.14 & 2.595 & 34499.6 &  22.524 & 5.48 & 11189 & 16.52 & 2.076&19.437&0.433 & 0.603&0.654 & S\\
MWISP G115.395$+$04.094$-$023.10 & 115.395&4.094&-23.10 & 2.057 & 17800.6 &  17.411 & 2.16 & 10640 & 9.63 & 0.748&5.236&0.532 & 0.888&1.073 & M\\
MWISP G106.100$+$00.590$-$002.94 & 106.100&0.590&-2.94 & 0.678 & 1561.5 &  5.688 & 2.13 & 10453 & 5.78 & 0.247&5.374&0.415 & 0.780&1.080 & S\\
MWISP G145.416$+$00.871$-$032.88 & 145.416&0.871&-32.88 & 3.044 & 33787.6 &  25.515 & 6.77 & 10434 & 17.83 & 1.107&15.717&0.548 & 0.818&0.876 & S\\
MWISP G027.893$-$02.170$+$018.03 & 27.893&-2.170&18.03 & 1.202 & 3485.3 &  9.969 & 5.86 & 10216 & 18.40 & 0.437&2.360&0.673 & 1.078&1.226 & M\\
MWISP G145.607$-$00.377$-$009.04 & 145.607&-0.377&-9.04 & 0.822 & 2470.9 &  6.546 & 4.43 & 9419 & 11.51 & 0.658&3.646&0.350 & 0.602&0.784 & M\\
MWISP G112.162$-$02.435$-$001.23 & 112.162&-2.435&-1.23 & 0.390 & 179.8 &  3.066 & 12.10 & 9177 & 3.56 & 0.142&2.865&0.408 & 0.797&1.176 & S\\
MWISP G040.729$-$04.115$+$007.83 & 40.729&-4.115&7.83 & 0.384 & 208.1 &  3.007 & 4.96 & 9104 & 2.31 & 0.140&2.094&0.446 & 0.825&1.137 & S\\
MWISP G209.003$+$02.193$+$009.67 & 209.003&2.193&9.67 & 0.928 & 1980.7 &  6.993 & 4.11 & 8434 & 14.30 & 0.337&6.276&0.446 & 0.730&0.860 & S\\
MWISP G148.160$-$00.237$-$004.18 & 148.160&-0.237&-4.18 & 0.458 & 849.5 &  3.255 & 13.34 & 7500 & 12.85 & 0.167&2.232&0.682 & 1.125&1.364 & S\\
MWISP G148.541$-$00.088$-$034.08 & 148.541&-0.088&-34.08 & 3.447 & 41482.3 &  24.414 & 0.56 & 7450 & 10.22 & 1.253&16.795&0.574 & 1.009&1.287 & S\\
MWISP G221.570$-$02.857$+$012.81 & 221.570&-2.857&12.81 & 1.038 & 3557.1 &  7.340 & 1.44 & 7426 & 9.97 & 0.377&3.850&0.482 & 0.771&0.873 & S\\
MWISP G123.291$-$00.539$-$044.15 & 123.291&-0.539&-44.15 & 3.522 & 33705.6 &  24.870 & 2.45 & 7405 & 12.07 & 1.281&8.964&0.630 & 1.062&1.271 & M\\
MWISP G220.607$-$01.924$+$011.77 & 220.607&-1.924&11.77 & 0.964 & 2927.7 &  6.605 & 2.60 & 6971 & 5.23 & 0.351&2.454&0.541 & 0.942&1.199 & M\\
MWISP G196.292$+$00.336$+$004.84 & 196.292&0.336&4.84 & 0.733 & 781.9 &  4.741 & 7.74 & 6213 & 5.60 & 0.267&4.638&0.628 & 1.113&1.452 & S\\
MWISP G136.806$+$01.256$-$038.05 & 136.806&1.256&-38.05 & 3.143 & 71661.2 &  19.921 & 1.10 & 5966 & 11.75 & ...&...&... & ...&... & N\\
MWISP G148.914$+$02.640$-$003.35 & 148.914&2.640&-3.35 & 0.397 & 229.2 &  2.494 & 2.20 & 5860 & 5.91 & 0.144&2.339&0.396 & 0.711&0.957 & S\\
MWISP G143.021$+$00.784$-$000.51 & 143.021&0.784&-0.51 & 0.179 & 24.2 &  1.036 & 29.91 & 4971 & 8.76 & 0.065&0.742&0.495 & 0.833&1.054 & S\\
MWISP G147.461$-$04.017$-$024.86 & 147.461&-4.017&-24.86 & 2.316 & 4913.2 &  13.146 & 6.32 & 4785 & 3.43 & 0.842&11.958&0.511 & 0.899&1.201 & S\\
MWISP G130.143$+$00.594$-$011.52 & 130.143&0.594&-11.52 & 0.962 & 1452.2 &  5.343 & 1.94 & 4582 & 3.56 & 0.350&1.469&0.489 & 0.869&1.114 & M\\
MWISP G206.477$+$02.119$+$013.33 & 206.477&2.119&13.33 & 1.354 & 1061.2 &  7.461 & 8.36 & 4509 & 3.44 & 0.492&7.385&0.577 & 1.080&1.516 & S\\
MWISP G048.907$+$02.291$+$006.87 & 48.907&2.291&6.87 & 0.301 & 68.1 &  1.612 & 1.44 & 4259 & 0.96 & 0.109&0.722&0.345 & 0.643&0.894 & S\\
MWISP G138.735$+$01.438$-$039.07 & 138.735&1.438&-39.07 & 3.321 & 28745.3 &  17.562 & 0.42 & 4153 & 2.26 & 1.208&16.181&0.464 & 0.867&1.202 & S\\
MWISP G144.096$+$04.220$-$009.89 & 144.096&4.220&-9.89 & 0.879 & 483.0 &  4.615 & 21.61 & 4093 & 8.07 & ...&...&... & ...&... & N\\
MWISP G141.520$-$03.424$-$018.08 & 141.520&-3.424&-18.08 & 1.502 & 1401.5 &  7.826 & 4.95 & 4032 & 8.94 & 0.546&6.008&0.557 & 1.058&1.490 & S\\
MWISP G108.614$-$01.006$-$051.24 & 108.614&-1.006&-51.24 & 4.616 & 120669.2 &  23.761 & 0.40 & 3935 & 3.76 & ...&...&... & ...&... & N\\
MWISP G217.803$-$00.246$+$027.03 & 217.803&-0.246&27.03 & 2.267 & 22603.1 &  11.329 & 1.11 & 3709 & 7.72 & 0.824&6.430&0.679 & 1.150&1.435 & S\\
MWISP G129.054$-$00.099$-$035.23 & 129.054&-0.099&-35.23 & 2.760 & 9144.2 &  13.270 & 5.92 & 3433 & 9.35 & ...&...&... & ...&... & N\\
MWISP G105.373$+$00.280$-$052.47 & 105.373&0.280&-52.47 & 4.936 & 35048.3 &  23.722 & 3.14 & 3430 & 17.67 & ...&...&... & ...&... & N\\
MWISP G206.698$-$04.407$+$009.26 & 206.698&-4.407&9.26 & 0.942 & 1013.1 &  4.425 & 8.49 & 3277 & 5.23 & 0.343&1.987&0.539 & 0.949&1.219 & M\\
MWISP G115.667$-$01.608$-$040.66 & 115.667&-1.608&-40.66 & 3.409 & 30019.7 &  15.819 & 1.68 & 3198 & 26.31 & 1.240&6.198&0.687 & 1.171&1.388 & M\\
MWISP G214.689$-$01.865$+$027.63 & 214.689&-1.865&27.63 & 2.453 & 8593.3 &  11.276 & 0.54 & 3138 & 2.20 & ...&...&... & ...&... & N\\
MWISP G142.433$-$00.761$-$015.35 & 142.433&-0.761&-15.35 & 1.289 & 784.6 &  5.872 & 13.89 & 3082 & 11.05 & 0.469&4.218&0.630 & 1.079&1.302 & S\\
MWISP G140.605$+$00.378$-$038.24 & 140.605&0.378&-38.24 & 3.327 & 20461.0 &  15.141 & 0.94 & 3076 & 2.49 & 1.210&4.597&0.679 & 1.205&1.613 & M\\
MWISP G149.535$-$01.088$-$007.65 & 149.535&-1.088&-7.65 & 0.753 & 1031.0 &  3.406 & 5.93 & 3038 & 4.50 & 0.274&3.340&0.465 & 0.853&1.188 & S\\
MWISP G041.981$+$02.300$+$005.37 & 41.981&2.300&5.37 & 0.219 & 20.3 &  0.976 & 11.47 & 2951 & 20.35 & 0.080&0.812&0.453 & 0.845&0.995 & S\\
MWISP G211.671$+$02.298$+$007.35 & 211.671&2.298&7.35 & 0.691 & 256.0 &  3.073 & 9.48 & 2938 & 18.07 & ...&...&... & ...&... & N\\
MWISP G120.698$+$01.167$-$062.12 & 120.698&1.167&-62.12 & 5.177 & 21098.4 &  22.916 & 5.19 & 2910 & 26.20 & 1.882&14.683&0.743 & 1.245&1.446 & S\\
MWISP G208.863$-$02.722$+$024.02 & 208.863&-2.722&24.02 & 2.408 & 3443.1 &  10.214 & 3.46 & 2672 & 9.33 & 0.876&8.931&0.572 & 0.987&1.221 & S\\
MWISP G221.933$-$02.107$+$039.51 & 221.933&-2.107&39.51 & 3.312 & 25427.5 &  13.922 & 0.34 & 2624 & 1.93 & 1.204&6.021&0.432 & 0.808&1.142 & M\\
MWISP G113.289$-$00.732$-$038.49 & 113.289&-0.732&-38.49 & 3.327 & 10419.4 &  13.966 & 32.67 & 2617 & 14.58 & 1.210&7.016&0.757 & 1.087&1.215 & S\\
MWISP G116.791$-$03.157$-$036.79 & 116.791&-3.157&-36.79 & 3.069 & 5923.9 &  12.787 & 1.86 & 2578 & 8.38 & 1.116&10.936&0.537 & 1.001&1.329 & S\\
MWISP G106.660$+$01.012$-$011.68 & 106.660&1.012&-11.68 & 1.479 & 3356.1 &  6.138 & 1.33 & 2558 & 12.17 & 0.968&3.549&0.421 & 0.684&0.822 & M\\
MWISP G147.589$-$01.153$-$000.54 & 147.589&-1.153&-0.54 & 0.178 & 12.8 &  0.737 & 26.91 & 2543 & 6.17 & 0.065&0.712&0.529 & 0.892&1.155 & S\\
MWISP G029.851$-$04.537$+$017.40 & 29.851&-4.537&17.40 & 1.120 & 763.5 &  4.608 & 1.98 & 2514 & 4.10 & 0.407&3.014&0.546 & 1.040&1.435 & S\\
MWISP G143.707$-$03.342$-$035.25 & 143.707&-3.342&-35.25 & 3.204 & 9085.0 &  12.826 & 0.75 & 2380 & 3.05 & 1.165&5.359&0.570 & 0.967&1.208 & M\\
MWISP G114.783$+$00.817$+$002.24 & 114.783&0.817&2.24 & 0.066 & 2.1 &  0.262 & 147.59 & 2343 & 3.66 & ...&...&... & ...&... & N\\
MWISP G217.055$+$00.622$+$050.10 & 217.055&0.622&50.10 & 4.917 & 12043.6 &  19.388 & 4.86 & 2309 & 8.33 & 1.788&8.224&0.845 & 1.353&1.599 & M\\
MWISP G143.542$-$01.598$-$014.85 & 143.542&-1.598&-14.85 & 1.265 & 581.9 &  4.953 & 20.37 & 2277 & 4.03 & 0.460&2.484&0.684 & 1.207&1.584 & S\\
MWISP G143.202$+$03.009$-$010.36 & 143.202&3.009&-10.36 & 0.906 & 404.0 &  3.391 & 8.68 & 2080 & 4.31 & ...&...&... & ...&... & N\\
MWISP G214.979$+$00.850$+$047.55 & 214.979&0.850&47.55 & 4.825 & 17601.9 &  17.870 & 1.36 & 2037 & 7.07 & ...&...&... & ...&... & N\\
MWISP G209.566$-$00.565$+$012.12 & 209.566&-0.565&12.12 & 1.139 & 519.9 &  4.214 & 8.49 & 2033 & 7.69 & 0.414&3.065&0.588 & 0.966&1.128 & S\\
MWISP G132.035$-$01.123$-$014.04 & 132.035&-1.123&-14.04 & 1.135 & 615.0 &  4.191 & 4.90 & 2025 & 12.09 & 0.413&1.898&0.649 & 1.175&1.508 & M\\
MWISP G210.544$-$03.342$+$019.58 & 210.544&-3.342&19.58 & 1.833 & 1063.2 &  6.767 & 9.85 & 2024 & 10.60 & ...&...&... & ...&... & N\\
MWISP G135.119$-$00.436$-$041.64 & 135.119&-0.436&-41.64 & 3.429 & 4128.1 &  12.384 & 5.40 & 1937 & 5.57 & 1.247&7.232&0.589 & 0.987&1.202 & S\\
MWISP G196.669$+$00.683$+$020.14 & 196.669&0.683&20.14 & 3.445 & 5128.0 &  12.172 & 2.46 & 1854 & 5.69 & 1.253&6.764&0.562 & 0.988&1.256 & S\\
MWISP G120.997$-$01.446$-$019.47 & 120.997&-1.446&-19.47 & 1.634 & 1037.0 &  5.730 & 6.34 & 1826 & 10.30 & 1.069&3.446&0.536 & 1.162&1.761 & M\\
MWISP G110.768$+$03.849$-$004.82 & 110.768&3.849&-4.82 & 0.733 & 291.8 &  2.563 & 16.97 & 1816 & 3.47 & ...&...&... & ...&... & N\\
MWISP G120.763$-$00.355$-$046.99 & 120.763&-0.355&-46.99 & 3.796 & 6074.6 &  13.120 & 2.38 & 1774 & 3.08 & 1.380&10.214&0.601 & 1.061&1.397 & S\\
MWISP G111.462$+$02.411$-$051.32 & 111.462&2.411&-51.32 & 4.473 & 11106.4 &  15.337 & 8.43 & 1746 & 14.16 & 1.626&8.132&0.742 & 1.343&1.752 & S\\
MWISP G117.414$+$02.498$-$005.10 & 117.414&2.498&-5.10 & 0.631 & 153.0 &  2.133 & 15.38 & 1697 & 4.70 & 0.229&2.065&0.631 & 1.176&1.657 & S\\
MWISP G222.290$-$04.709$+$016.40 & 222.290&-4.709&16.40 & 1.310 & 675.6 &  4.396 & 5.48 & 1672 & 2.97 & 0.476&2.191&0.610 & 1.199&1.765 & M\\
MWISP G196.590$-$01.527$+$015.67 & 196.590&-1.527&15.67 & 2.510 & 6227.3 &  8.412 & 2.31 & 1668 & 1.82 & 0.913&8.214&0.458 & 0.889&1.279 & S\\
MWISP G041.581$+$02.328$+$017.64 & 41.581&2.328&17.64 & 0.994 & 262.1 &  3.302 & 17.13 & 1639 & 7.29 & 0.361&1.807&0.429 & 0.865&1.231 & S\\
MWISP G106.065$+$01.491$-$023.46 & 106.065&1.491&-23.46 & 2.510 & 1945.0 &  8.318 & 5.32 & 1631 & 10.02 & 0.913&8.214&0.648 & 1.073&1.321 & S\\
MWISP G030.276$-$01.657$+$023.26 & 30.276&-1.657&23.26 & 1.471 & 550.9 &  4.872 & 9.86 & 1629 & 2.45 & 0.535&3.744&0.638 & 1.176&1.610 & S\\
MWISP G130.866$-$01.000$-$010.29 & 130.866&-1.000&-10.29 & 0.873 & 247.7 &  2.885 & 2.77 & 1622 & 1.99 & 0.317&2.857&0.408 & 0.718&0.940 & S\\
MWISP G133.505$-$01.634$+$001.06 & 133.505&-1.634&1.06 & 0.093 & 2.9 &  0.306 & 53.09 & 1611 & 4.58 & 0.034&0.237&0.572 & 0.973&1.226 & S\\
MWISP G141.548$-$01.197$-$041.03 & 141.548&-1.197&-41.03 & 3.690 & 12247.2 &  12.150 & 0.66 & 1610 & 4.85 & 1.342&8.319&0.629 & 1.168&1.588 & S\\
MWISP G202.854$+$01.263$+$031.72 & 202.854&1.263&31.72 & 4.384 & 5170.7 &  14.412 & 9.30 & 1605 & 6.03 & 1.594&9.883&0.610 & 0.997&1.243 & S\\
MWISP G107.866$+$01.907$-$011.13 & 107.866&1.907&-11.13 & 1.377 & 621.3 &  4.489 & 17.53 & 1578 & 5.25 & 0.501&2.503&0.588 & 1.131&1.604 & S\\
MWISP G146.013$-$04.267$-$009.16 & 146.013&-4.267&-9.16 & 0.841 & 166.9 &  2.741 & 7.60 & 1578 & 6.75 & 0.550&2.018&0.718 & 1.322&1.801 & M\\
MWISP G130.454$+$04.799$-$014.05 & 130.454&4.799&-14.05 & 1.148 & 379.0 &  3.740 & 11.67 & 1576 & 1.86 & 0.417&1.753&0.638 & 1.203&1.712 & M\\
MWISP G129.945$-$02.148$-$013.46 & 129.945&-2.148&-13.46 & 1.102 & 281.1 &  3.544 & 2.83 & 1536 & 0.68 & 0.401&1.683&0.467 & 0.954&1.450 & M\\
MWISP G114.581$-$00.421$-$049.35 & 114.581&-0.421&-49.35 & 4.163 & 22824.4 &  13.301 & 0.50 & 1516 & 3.98 & 1.514&5.752&0.530 & 0.996&1.407 & M\\
MWISP G111.660$+$04.099$-$030.30 & 111.660&4.099&-30.30 & 2.766 & 1752.5 &  8.811 & 4.84 & 1507 & 2.76 & 1.006&8.649&0.516 & 0.975&1.385 & S\\
MWISP G140.855$-$01.152$-$040.27 & 140.855&-1.152&-40.27 & 3.560 & 18198.1 &  11.306 & 0.88 & 1498 & 7.88 & 1.294&9.061&0.629 & 1.147&1.497 & S\\
MWISP G123.438$+$03.213$-$061.52 & 123.438&3.213&-61.52 & 5.138 & 10576.3 &  16.225 & 2.92 & 1481 & 3.44 & 1.868&13.078&0.714 & 1.281&1.722 & S\\
MWISP G131.838$+$00.093$-$015.56 & 131.838&0.093&-15.56 & 1.244 & 491.4 &  3.891 & 3.63 & 1453 & 2.21 & 0.452&1.538&0.500 & 0.926&1.261 & M\\
MWISP G104.912$+$02.756$-$024.20 & 104.912&2.756&-24.20 & 2.649 & 2507.5 &  8.223 & 0.84 & 1431 & 1.73 & 0.963&7.513&0.576 & 1.082&1.506 & S\\
MWISP G208.887$+$01.874$+$016.12 & 208.887&1.874&16.12 & 1.547 & 341.3 &  4.740 & 16.00 & 1394 & 8.32 & ...&...&... & ...&... & N\\
MWISP G212.966$+$01.223$+$042.73 & 212.966&1.223&42.73 & 4.406 & 9158.6 &  13.470 & 1.65 & 1388 & 2.25 & 1.602&12.496&0.668 & 1.221&1.663 & S\\
MWISP G146.134$-$00.247$-$032.19 & 146.134&-0.247&-32.19 & 3.016 & 3399.3 &  9.207 & 1.12 & 1384 & 3.41 & 1.097&5.045&0.571 & 1.017&1.297 & S\\
MWISP G137.708$+$01.499$-$039.27 & 137.708&1.499&-39.27 & 3.297 & 23077.6 &  9.999 & 0.89 & 1366 & 5.47 & 1.199&7.912&0.610 & 1.091&1.457 & S\\
MWISP G117.004$-$02.268$-$043.15 & 117.004&-2.268&-43.15 & 3.570 & 5836.6 &  10.772 & 5.09 & 1352 & 12.77 & 1.298&10.644&0.540 & 0.929&1.111 & S\\
MWISP G224.530$-$02.548$+$012.35 & 224.530&-2.548&12.35 & 0.996 & 1786.1 &  2.975 & 2.43 & 1325 & 5.76 & 0.362&2.535&0.639 & 1.196&1.656 & S\\
MWISP G038.082$-$02.627$+$007.23 & 38.082&-2.627&7.23 & 0.357 & 15.6 &  1.046 & 6.48 & 1274 & 2.76 & ...&...&... & ...&... & N\\
MWISP G145.360$-$03.258$-$019.25 & 145.360&-3.258&-19.25 & 1.685 & 660.0 &  4.933 & 2.67 & 1273 & 3.29 & 0.613&2.328&0.549 & 0.934&1.203 & M\\
MWISP G131.993$-$00.818$-$054.23 & 131.993&-0.818&-54.23 & 4.587 & 7273.4 &  13.377 & 2.05 & 1263 & 4.26 & ...&...&... & ...&... & N\\
MWISP G117.912$-$03.254$-$038.50 & 117.912&-3.254&-38.50 & 3.172 & 3827.7 &  9.225 & 6.74 & 1256 & 5.00 & 1.153&8.996&0.702 & 1.298&1.724 & S\\
MWISP G109.693$+$01.885$-$057.12 & 109.693&1.885&-57.12 & 5.062 & 8178.6 &  14.562 & 5.46 & 1229 & 1.92 & 1.841&14.357&0.634 & 1.148&1.559 & S\\
MWISP G224.370$-$03.770$+$015.18 & 224.370&-3.770&15.18 & 1.205 & 833.0 &  3.448 & 0.47 & 1216 & 1.73 & 0.438&3.418&0.396 & 0.735&1.011 & S\\
MWISP G146.273$-$02.388$-$010.51 & 146.273&-2.388&-10.51 & 0.950 & 168.8 &  2.699 & 29.51 & 1199 & 0.93 & 0.345&1.036&0.627 & 1.178&1.685 & M\\
MWISP G029.340$-$04.870$+$010.56 & 29.340&-4.870&10.56 & 0.671 & 143.9 &  1.879 & 1.41 & 1164 & 3.93 & 0.244&1.220&0.555 & 1.023&1.385 & S\\
MWISP G223.750$-$04.108$+$011.87 & 223.750&-4.108&11.87 & 0.965 & 252.5 &  2.672 & 1.29 & 1139 & 2.28 & 0.351&2.175&0.452 & 0.822&1.102 & S\\
MWISP G148.423$-$01.420$-$025.49 & 148.423&-1.420&-25.49 & 2.426 & 2694.7 &  6.713 & 0.60 & 1137 & 1.08 & 0.882&3.705&0.548 & 1.041&1.497 & S\\
MWISP G117.486$-$00.610$+$000.68 & 117.486&-0.610&0.68 & 0.180 & 7.3 &  0.497 & 27.64 & 1130 & 1.67 & 0.065&0.484&0.520 & 0.984&1.374 & S\\
MWISP G138.475$-$00.745$-$045.63 & 138.475&-0.745&-45.63 & 3.998 & 4817.7 &  10.965 & 0.87 & 1117 & 5.13 & 1.454&4.943&0.598 & 1.024&1.247 & M\\
MWISP G107.305$+$00.037$-$050.61 & 107.305&0.037&-50.61 & 4.644 & 6840.9 &  12.650 & 3.31 & 1102 & 5.26 & 1.689&7.092&0.667 & 1.144&1.545 & S\\
MWISP G150.102$+$03.572$-$008.74 & 150.102&3.572&-8.74 & 0.858 & 322.2 &  2.336 & 2.98 & 1101 & 1.05 & 0.312&1.685&0.549 & 1.041&1.469 & S\\
MWISP G212.141$-$01.044$+$044.18 & 212.141&-1.044&44.18 & 4.739 & 9633.8 &  12.880 & 2.20 & 1097 & 10.07 & ...&...&... & ...&... & N\\
MWISP G142.726$-$00.972$-$040.66 & 142.726&-0.972&-40.66 & 3.733 & 2440.1 &  9.964 & 0.93 & 1058 & 1.93 & 1.357&9.501&0.517 & 0.926&1.235 & S\\
MWISP G035.767$-$04.960$+$008.81 & 35.767&-4.960&8.81 & 0.479 & 29.6 &  1.273 & 13.68 & 1049 & 13.77 & 0.174&0.732&0.567 & 1.177&1.564 & S\\
MWISP G130.587$+$01.821$-$044.36 & 130.587&1.821&-44.36 & 3.573 & 3556.3 &  9.455 & 1.38 & 1040 & 5.21 & ...&...&... & ...&... & N\\
MWISP G204.790$+$00.474$+$009.58 & 204.790&0.474&9.58 & 1.019 & 258.4 &  2.697 & 1.99 & 1040 & 2.17 & 0.667&1.853&0.338 & 0.609&0.818 & M\\
MWISP G211.994$+$04.394$+$006.70 & 211.994&4.394&6.70 & 0.638 & 50.3 &  1.686 & 30.58 & 1037 & 7.33 & 0.232&0.789&0.724 & 1.271&1.600 & M\\
MWISP G106.493$+$01.045$-$061.01 & 106.493&1.045&-61.01 & 5.600 & 8614.4 &  14.769 & 1.25 & 1033 & 3.75 & 2.851&14.254&0.429 & 0.848&1.206 & S\\
MWISP G144.078$-$03.824$-$007.37 & 144.078&-3.824&-7.37 & 0.684 & 46.7 &  1.800 & 115.57 & 1029 & 4.67 & ...&...&... & ...&... & N\\
MWISP G149.139$-$02.111$-$008.78 & 149.139&-2.111&-8.78 & 0.845 & 782.3 &  2.215 & 1.51 & 1020 & 1.79 & 0.307&2.151&0.649 & 1.169&1.589 & S\\
MWISP G207.823$+$01.124$+$016.92 & 207.823&1.124&16.92 & 1.674 & 523.9 &  4.381 & 6.48 & 1017 & 6.14 & 0.609&2.313&0.695 & 1.205&1.599 & M\\
MWISP G139.478$+$02.917$-$051.28 & 139.478&2.917&-51.28 & 4.752 & 4511.8 &  12.435 & 4.41 & 1017 & 6.38 & 1.728&7.948&0.803 & 1.528&2.055 & S\\
MWISP G133.440$-$02.068$-$012.63 & 133.440&-2.068&-12.63 & 1.033 & 163.3 &  2.671 & 19.89 & 993 & 15.73 & 0.376&1.728&0.741 & 1.371&1.658 & S\\
MWISP G199.973$+$01.326$+$020.78 & 199.973&1.326&20.78 & 2.902 & 2883.9 &  7.500 & 0.89 & 992 & 2.83 & 1.055&3.588&0.645 & 1.314&1.914 & M\\
MWISP G119.262$-$01.621$-$034.08 & 119.262&-1.621&-34.08 & 2.780 & 2563.6 &  7.083 & 5.37 & 964 & 2.40 & ...&...&... & ...&... & N\\
MWISP G134.067$-$04.075$-$013.79 & 134.067&-4.075&-13.79 & 1.120 & 258.2 &  2.788 & 13.47 & 920 & 3.41 & 0.407&2.688&0.809 & 1.420&1.878 & S\\
MWISP G049.197$-$04.099$+$009.87 & 49.197&-4.099&9.87 & 0.500 & 57.9 &  1.243 & 0.78 & 918 & 1.70 & 0.182&0.691&0.438 & 0.745&0.950 & M\\
MWISP G044.760$+$04.035$+$020.00 & 44.760&4.035&20.00 & 1.142 & 131.6 &  2.836 & 13.39 & 916 & 1.14 & 0.415&2.242&0.743 & 1.562&2.386 & S\\
MWISP G195.753$-$02.284$+$004.42 & 195.753&-2.284&4.42 & 0.693 & 345.7 &  1.719 & 0.69 & 914 & 2.25 & 0.353&1.663&0.415 & 0.747&1.021 & S\\
MWISP G215.242$-$00.376$+$028.93 & 215.242&-0.376&28.93 & 2.555 & 895.9 &  6.297 & 5.92 & 902 & 26.77 & ...&...&... & ...&... & N\\
MWISP G135.883$+$04.911$-$002.41 & 135.883&4.911&-2.41 & 0.320 & 12.7 &  0.786 & 22.39 & 897 & 11.26 & 0.116&0.768&0.654 & 0.991&1.100 & S\\
MWISP G109.514$+$03.624$-$010.25 & 109.514&3.624&-10.25 & 1.240 & 392.5 &  3.047 & 6.91 & 897 & 4.37 & 0.451&1.894&0.644 & 1.030&1.222 & S\\
MWISP G117.941$+$04.960$-$000.50 & 117.941&4.960&-0.50 & 0.270 & 11.2 &  0.661 & 9.97 & 891 & 1.71 & 0.137&0.648&0.577 & 1.028&1.387 & S\\
MWISP G040.524$+$02.087$+$008.20 & 40.524&2.087&8.20 & 0.407 & 19.6 &  0.997 & 13.81 & 891 & 2.94 & 0.148&0.681&0.583 & 1.010&1.286 & S\\
MWISP G210.327$-$00.047$+$036.13 & 210.327&-0.047&36.13 & 3.810 & 3732.1 &  9.316 & 1.35 & 888 & 2.29 & 1.385&9.143&0.377 & 0.725&1.044 & S\\
MWISP G205.866$+$00.227$+$019.93 & 205.866&0.227&19.93 & 2.130 & 840.1 &  5.120 & 1.06 & 858 & 0.88 & 0.774&2.014&0.530 & 0.968&1.314 & M\\
MWISP G112.149$+$05.107$-$010.95 & 112.149&5.107&-10.95 & 1.214 & 279.9 &  2.915 & 9.72 & 856 & 17.02 & 0.441&2.384&0.718 & 1.258&1.601 & S\\
MWISP G026.913$-$03.571$+$016.34 & 26.913&-3.571&16.34 & 1.118 & 210.3 &  2.681 & 1.99 & 854 & 2.42 & 0.407&1.870&0.623 & 1.180&1.648 & S\\
MWISP G114.203$-$01.009$-$032.51 & 114.203&-1.009&-32.51 & 2.817 & 2510.4 &  6.739 & 4.25 & 850 & 4.18 & 1.024&6.351&0.609 & 1.118&1.495 & S\\
MWISP G132.908$+$01.500$-$003.98 & 132.908&1.500&-3.98 & 0.432 & 16.0 &  1.022 & 69.92 & 831 & 17.23 & 0.157&0.974&0.993 & 1.633&1.911 & S\\
MWISP G139.725$-$00.507$-$038.33 & 139.725&-0.507&-38.33 & 3.291 & 1720.9 &  7.775 & 5.31 & 829 & 1.24 & 1.197&7.419&0.711 & 1.360&1.921 & S\\
MWISP G129.311$-$01.646$-$010.65 & 129.311&-1.646&-10.65 & 0.906 & 134.8 &  2.139 & 10.58 & 828 & 4.26 & 0.329&1.515&0.663 & 1.084&1.327 & S\\
MWISP G144.438$-$02.595$-$020.01 & 144.438&-2.595&-20.01 & 1.726 & 521.9 &  4.051 & 3.71 & 818 & 2.35 & ...&...&... & ...&... & N\\
MWISP G105.097$+$01.122$-$047.78 & 105.097&1.122&-47.78 & 4.564 & 6587.9 &  10.679 & 2.18 & 813 & 3.39 & 2.323&9.625&0.392 & 0.791&1.166 & S\\
MWISP G196.275$-$02.285$+$014.69 & 196.275&-2.285&14.69 & 2.372 & 1540.7 &  5.543 & 2.51 & 811 & 0.98 & 0.862&5.347&0.444 & 0.849&1.229 & S\\
MWISP G134.332$-$01.732$-$048.39 & 134.332&-1.732&-48.39 & 4.078 & 5606.2 &  9.530 & 0.60 & 811 & 1.64 & 1.483&6.228&0.482 & 0.919&1.285 & S\\
MWISP G195.552$-$00.136$+$020.02 & 195.552&-0.136&20.02 & 3.731 & 2949.7 &  8.665 & 0.56 & 801 & 3.68 & 1.357&5.155&0.476 & 0.809&0.978 & S\\
MWISP G224.890$+$00.737$+$017.38 & 224.890&0.737&17.38 & 1.362 & 371.8 &  3.143 & 0.60 & 791 & 1.58 & ...&...&... & ...&... & N\\
MWISP G120.494$+$01.828$-$062.22 & 120.494&1.828&-62.22 & 5.191 & 5718.6 &  11.973 & 1.06 & 790 & 3.15 & 1.887&6.417&0.634 & 1.128&1.472 & M\\
MWISP G206.244$-$00.741$+$010.46 & 206.244&-0.741&10.46 & 1.067 & 149.4 &  2.448 & 7.07 & 782 & 2.17 & 0.543&1.940&0.627 & 1.309&1.939 & M\\
MWISP G121.887$-$00.452$-$012.51 & 121.887&-0.452&-12.51 & 1.110 & 366.7 &  2.531 & 0.56 & 772 & 2.59 & 0.404&2.179&0.499 & 0.886&1.157 & S\\
MWISP G108.483$+$00.362$-$009.78 & 108.483&0.362&-9.78 & 1.234 & 242.3 &  2.812 & 8.03 & 771 & 1.73 & 0.628&1.885&0.574 & 1.140&1.719 & M\\
MWISP G209.565$+$00.605$+$030.11 & 209.565&0.605&30.11 & 3.088 & 1742.6 &  7.031 & 4.55 & 770 & 2.51 & 1.123&6.962&0.770 & 1.371&1.832 & S\\
MWISP G139.092$-$03.225$-$031.78 & 139.092&-3.225&-31.78 & 2.638 & 3779.6 &  5.968 & 0.44 & 760 & 1.33 & ...&...&... & ...&... & N\\
MWISP G118.684$+$02.959$-$066.59 & 118.684&2.959&-66.59 & 5.648 & 4859.4 &  12.777 & 9.05 & 760 & 16.56 & ... & ... & ... & ... & ... & N\\
MWISP G117.451$-$00.107$-$046.67 & 117.451&-0.107&-46.67 & 3.843 & 2717.2 &  8.676 & 2.82 & 757 & 3.75 & 1.397&3.633&0.755 & 1.376&1.794 & M\\
MWISP G114.573$-$00.108$-$036.22 & 114.573&-0.108&-36.22 & 3.093 & 2072.8 &  6.969 & 3.39 & 754 & 4.73 & 1.125&5.173&0.668 & 1.075&1.305 & S\\
MWISP G125.566$+$02.073$-$053.50 & 125.566&2.073&-53.50 & 4.361 & 8513.1 &  9.813 & 0.73 & 752 & 0.88 & 1.586&7.294&0.668 & 1.201&1.631 & S\\
MWISP G207.526$+$02.216$+$014.76 & 207.526&2.216&14.76 & 1.462 & 172.9 &  3.261 & 1.87 & 739 & 0.79 & 0.532&2.871&0.472 & 0.891&1.279 & S\\
MWISP G139.345$-$01.900$-$048.77 & 139.345&-1.900&-48.77 & 4.425 & 2903.4 &  9.797 & 1.85 & 728 & 1.33 & 1.609&6.114&0.594 & 1.068&1.455 & S\\
MWISP G197.015$-$03.153$+$023.88 & 197.015&-3.153&23.88 & 4.277 & 5129.5 &  9.450 & 2.61 & 725 & 1.50 & 1.555&7.154&0.703 & 1.369&1.968 & S\\
MWISP G197.653$-$03.026$+$003.82 & 197.653&-3.026&3.82 & 0.559 & 46.8 &  1.235 & 4.29 & 725 & 1.52 & 0.203&0.935&0.695 & 1.272&1.751 & S\\
MWISP G120.101$+$03.507$-$069.17 & 120.101&3.507&-69.17 & 5.914 & 17671.5 &  13.049 & 0.52 & 723 & 1.04 & 3.011&9.032&0.375 & 0.715&0.989 & M\\
MWISP G144.432$+$00.453$-$039.94 & 144.432&0.453&-39.94 & 3.789 & 2099.9 &  8.314 & 5.97 & 715 & 5.35 & 1.378&4.684&0.722 & 1.170&1.317 & M\\

		\enddata
		 \tablecomments{Column 1: name of the cloud. Columns 2-5: centroid Galactic longitude, latitude, velocity, and kinematic distance of the cloud. Columns 6-8: mass, effective radius, and virial parameter of the cloud. Column 9: pixel number. Column 10: kurtosis of velocity increments at the smallest lag. Columns 11-12: left end and right end of the power-law fitting range for the unweighted VSFs. Columns 13-15: fitted power-law exponents for the first to third-order unweighted VSFs. 
         Column 16: category of the unweighed VSF. The molecular clouds are listed in descending order according to the numbers of their projected pixels. The source name is defined under the MWISP standard. According to the MWISP standard for nomenclature, molecular clouds are named after their centroid positions and velocities. Specifically, the names start with ``MWISP" and then contain the spatial coordinates of the molecular clouds accurate to three decimal places and the centroid velocities accurate to two decimal places. The accuracy is set according to the pointing accuracy and the velocity resolution of the PMO-13.7m telescope.}
	\end{deluxetable*}
\end{longrotatetable}
\clearpage
\section{Supplementary Figures}

\begin{figure*}[htb!]
    \gridline{\fig{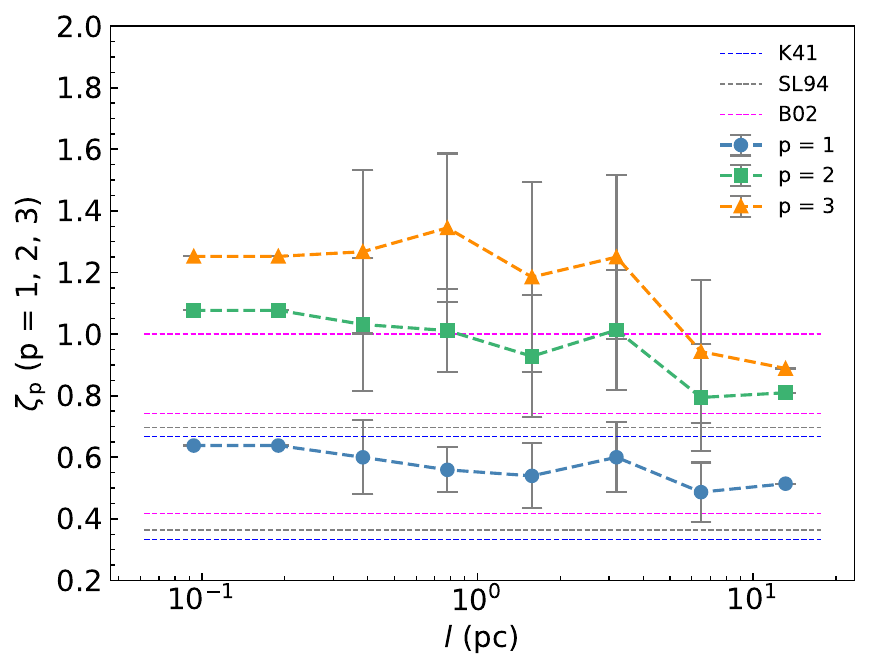}{0.49\textwidth}{(a)}
    \fig{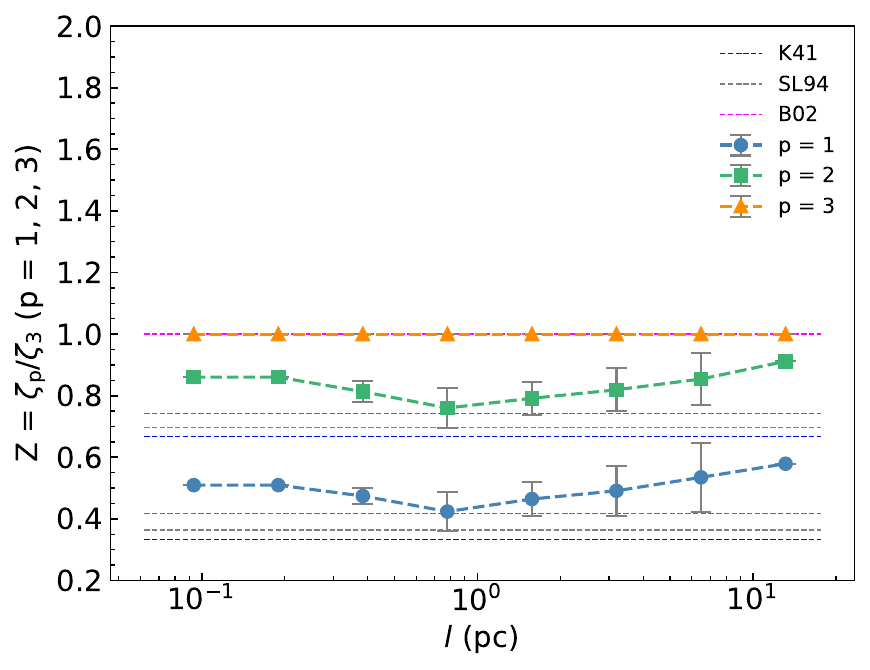}{0.49\textwidth}{(b)}}
    \caption{Same as Figure \ref{fig5}, but for the VSFs with only moderately significant power-law.}
    \label{groups_M}
\end{figure*}

\begin{figure*}[htb!]
    \gridline{\fig{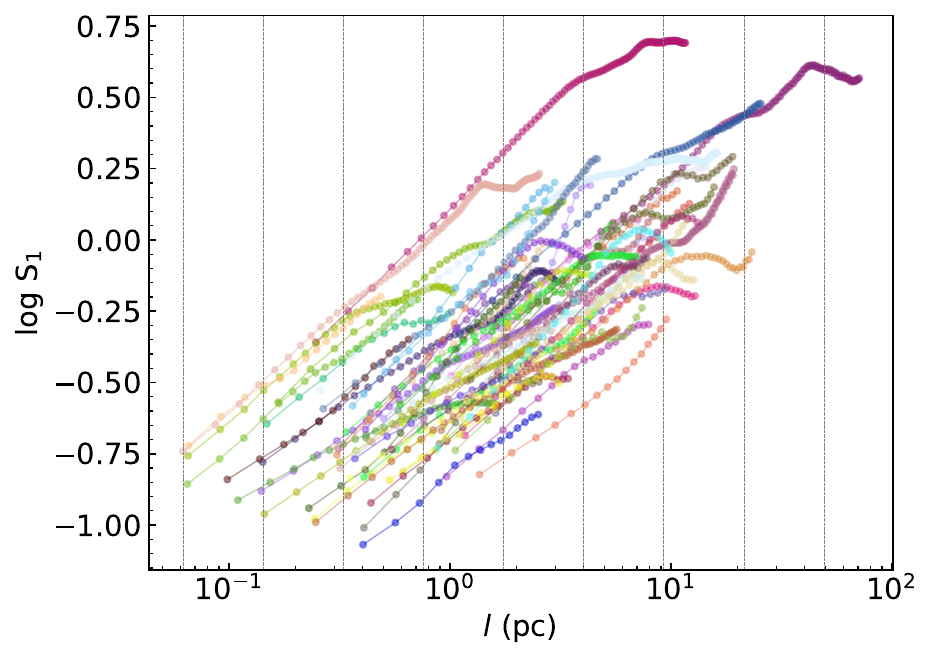}{0.33\textwidth}{(a)}
    \fig{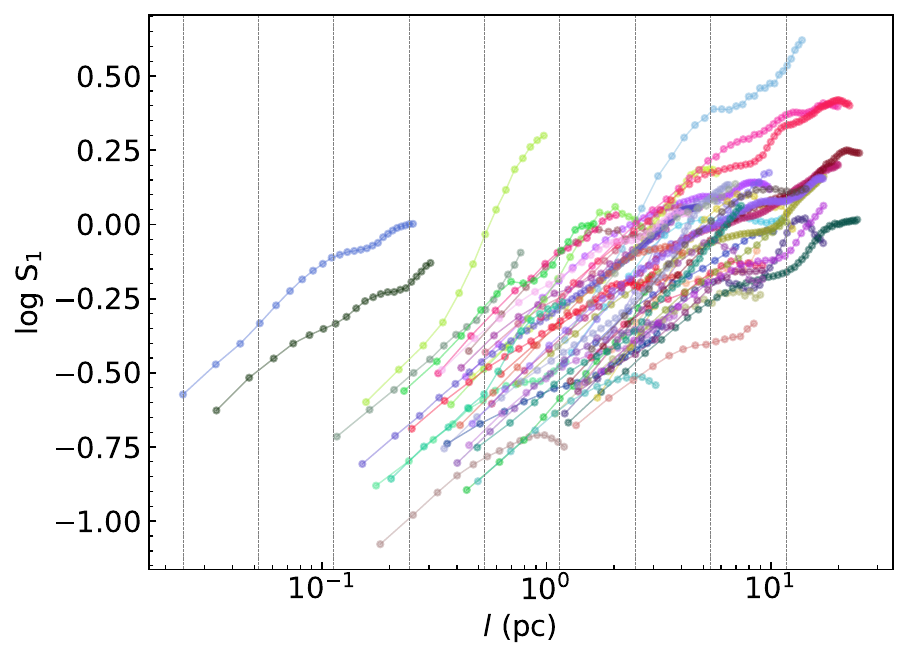}{0.33\textwidth}{(b)}
    \fig{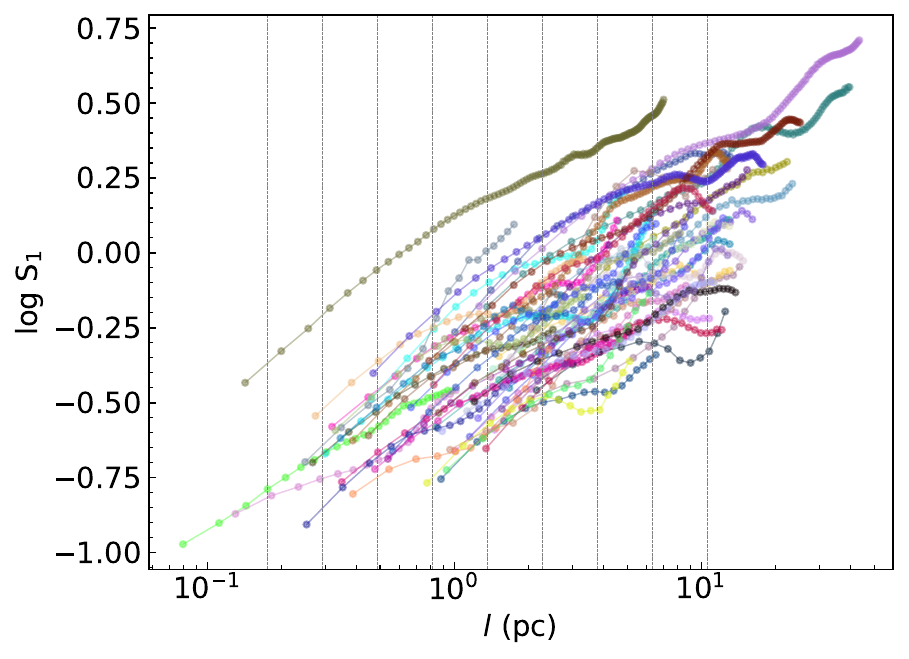}{0.33\textwidth}{(c)}}
    \gridline{\fig{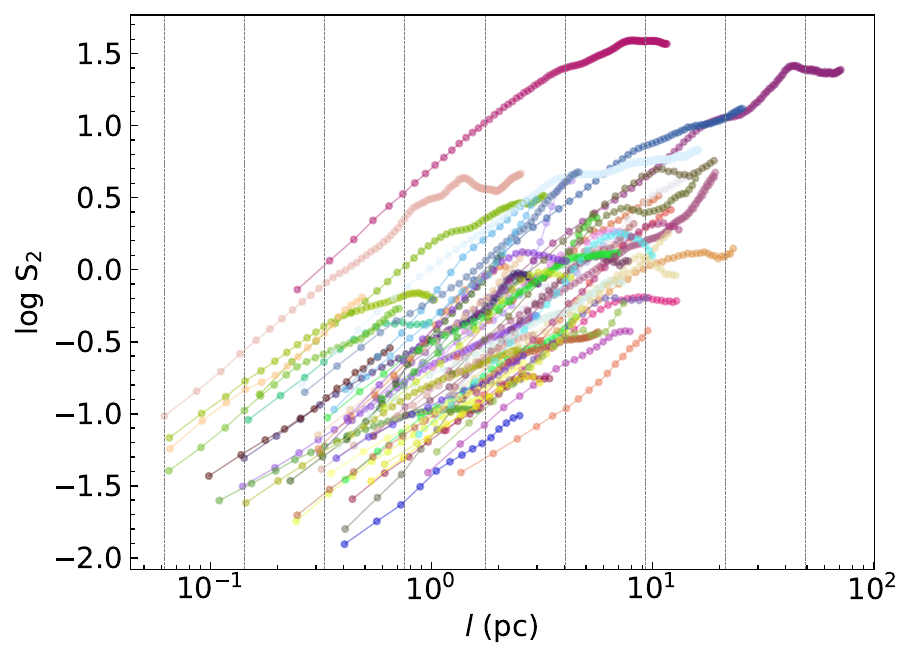}{0.33\textwidth}{(d)}
    \fig{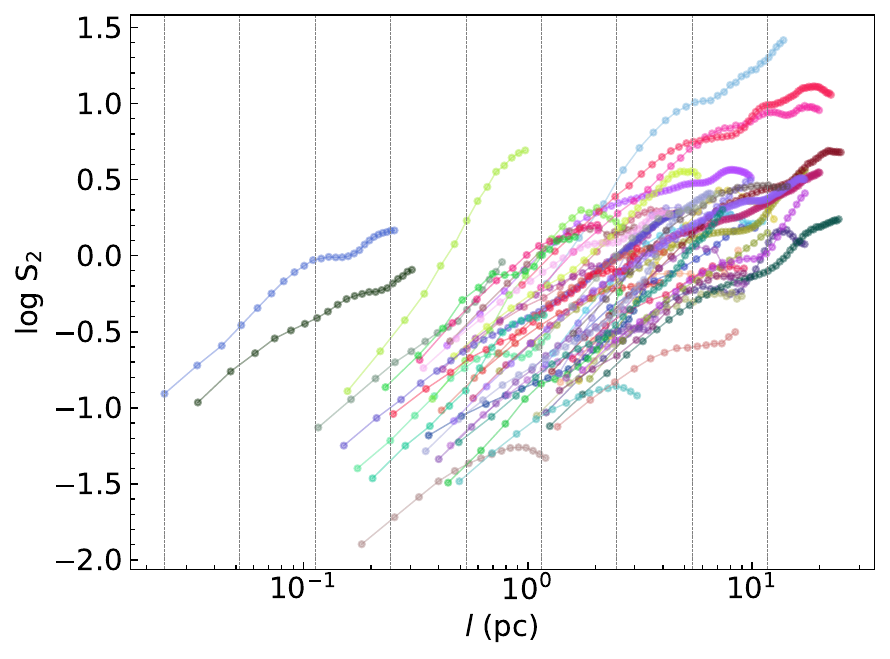}{0.33\textwidth}{(e)}
    \fig{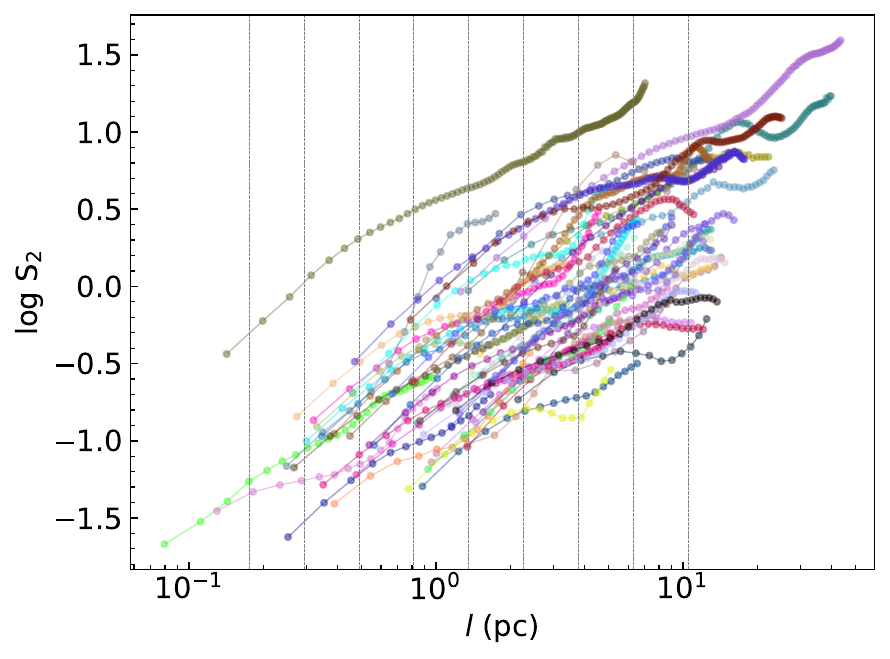}{0.33\textwidth}{(f)}}
    \gridline{\fig{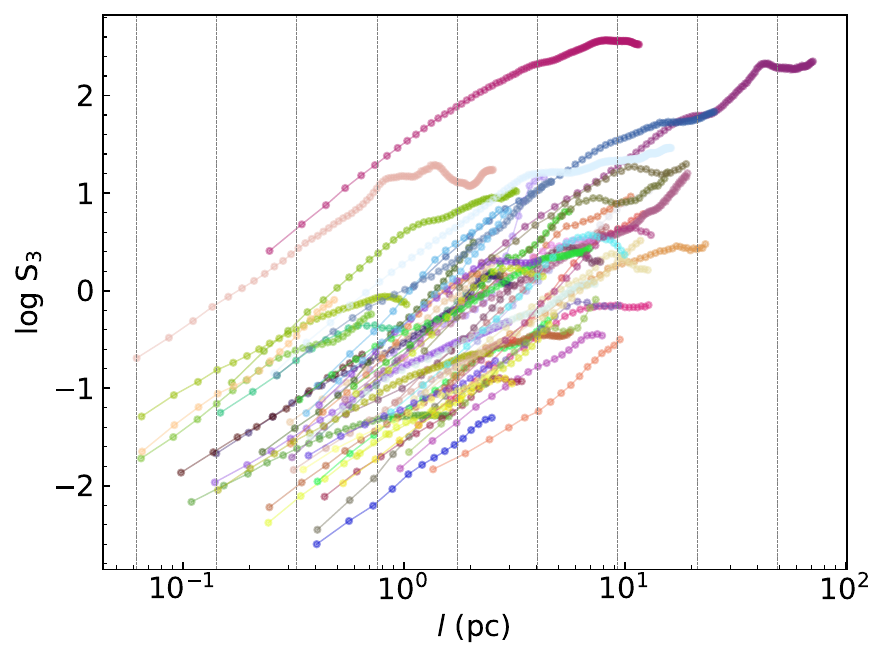}{0.33\textwidth}{(g)}
    \fig{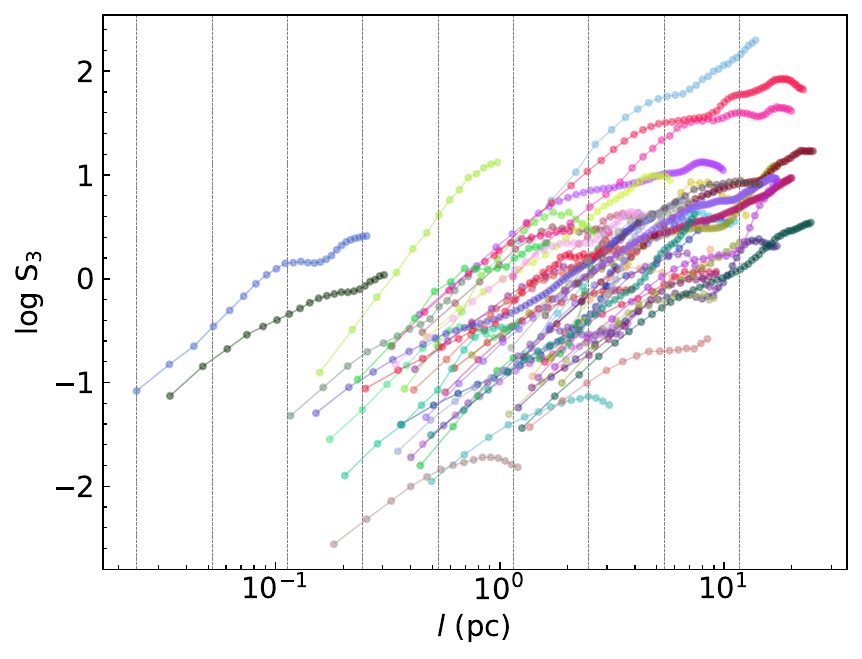}{0.33\textwidth}{(h)}
    \fig{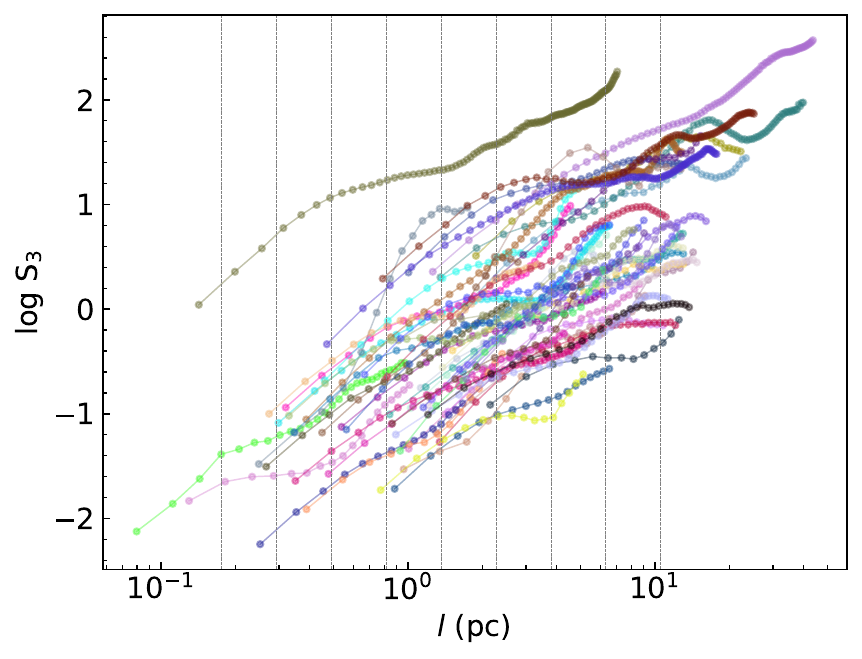}{0.33\textwidth}{(k)}}
    \caption{Same as Figure \ref{fig3}, but for the column density weighted VSFs.}
    \label{vsfs_wei}
\end{figure*}

\begin{figure*}[htb!]
    \gridline{\fig{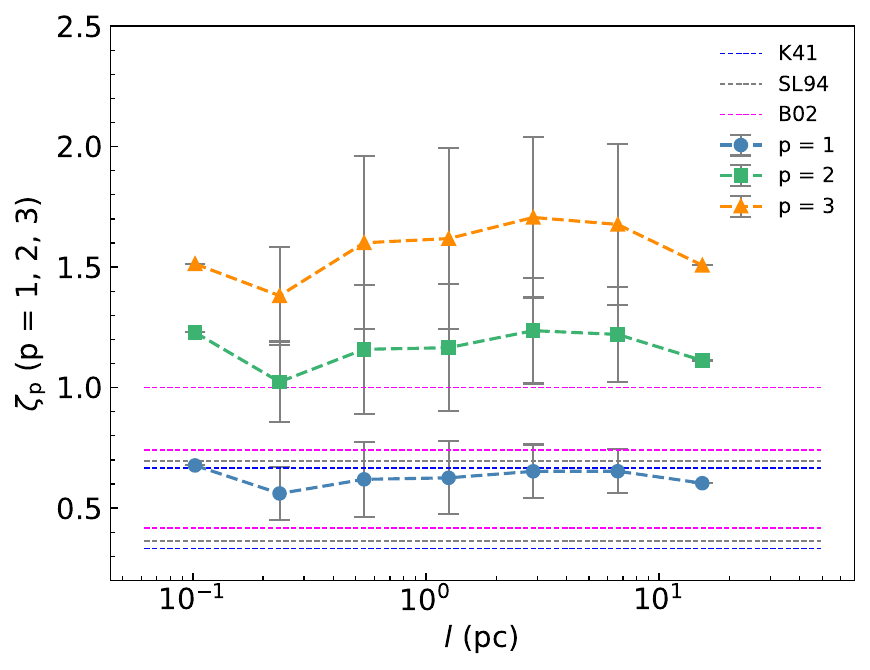}{0.49\textwidth}{(a)}
    \fig{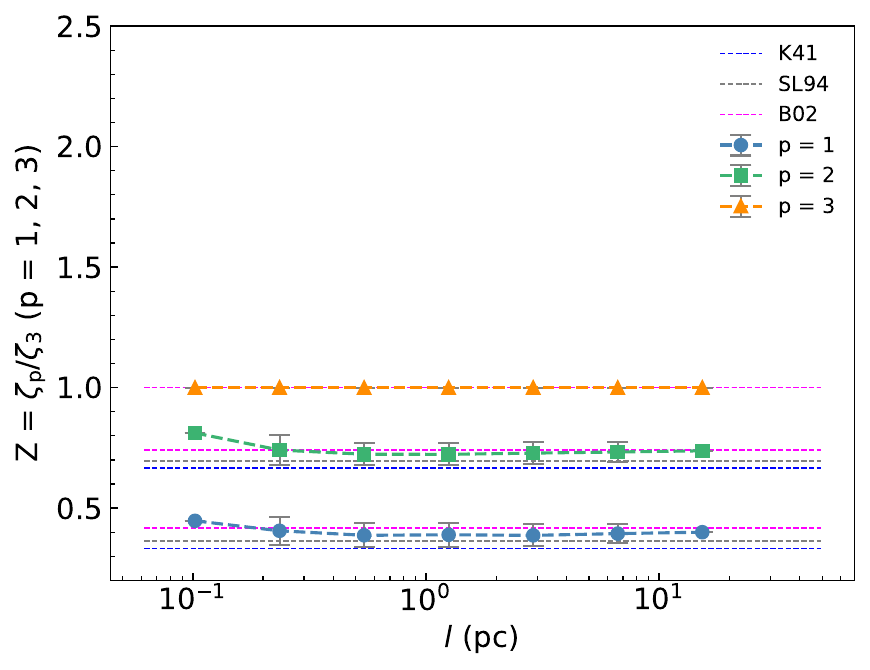}{0.49\textwidth}{(b)}}
    \caption{Same as Figure \ref{fig5}, but for the scaling exponents of the column density weighted VSFs.}
    \label{groups_S_wei}
\end{figure*}

% \section{Gold Open Access}

% \section{Author publication charges} \label{sec:pubcharge}

% \section{Rotating tables} \label{sec:rotate}

% \section{Using Chinese, Japanese, and Korean characters}

\clearpage
\bibliography{sample631.bbl}{}

\begin{thebibliography}{}
\expandafter\ifx\csname natexlab\endcsname\relax\def\natexlab#1{#1}\fi
\providecommand{\url}[1]{\href{#1}{#1}}
\providecommand{\dodoi}[1]{doi:~\href{http://doi.org/#1}{\nolinkurl{#1}}}
\providecommand{\doeprint}[1]{\href{http://ascl.net/#1}{\nolinkurl{http://ascl.net/#1}}}
\providecommand{\doarXiv}[1]{\href{https://arxiv.org/abs/#1}{\nolinkurl{https://arxiv.org/abs/#1}}}

\bibitem[{{Anselmet} {et~al.}(1984){Anselmet}, {Gagne}, {Hopfinger}, \&
  {Antonia}}]{Anselmet1984}
{Anselmet}, F., {Gagne}, Y., {Hopfinger}, E.~J., \& {Antonia}, R.~A. 1984,
  Journal of Fluid Mechanics, 140, 63, \dodoi{10.1017/S0022112084000513}

\bibitem[{{Ballesteros-Paredes} {et~al.}(2007){Ballesteros-Paredes}, {Klessen},
  {Mac Low}, \& {Vazquez-Semadeni}}]{Ballesteros-Paredes2007}
{Ballesteros-Paredes}, J., {Klessen}, R.~S., {Mac Low}, M.~M., \&
  {Vazquez-Semadeni}, E. 2007, in Protostars and Planets V, ed. B.~{Reipurth},
  D.~{Jewitt}, \& K.~{Keil}, 63, \dodoi{10.48550/arXiv.astro-ph/0603357}

\bibitem[{{Banerjee} {et~al.}(2009){Banerjee}, {V{\'a}zquez-Semadeni},
  {Hennebelle}, \& {Klessen}}]{Banerjee2009}
{Banerjee}, R., {V{\'a}zquez-Semadeni}, E., {Hennebelle}, P., \& {Klessen},
  R.~S. 2009, \mnras, 398, 1082, \dodoi{10.1111/j.1365-2966.2009.15115.x}

\bibitem[{{Batchelor} \& {Townsend}(1949)}]{Batchelor1949}
{Batchelor}, G.~K., \& {Townsend}, A.~A. 1949, Proceedings of the Royal Society
  of London Series A, 199, 238, \dodoi{10.1098/rspa.1949.0136}

\bibitem[{{Boldyrev}(2002)}]{Boldyrev2002a}
{Boldyrev}, S. 2002, \apj, 569, 841, \dodoi{10.1086/339403}

\bibitem[{{Boldyrev} {et~al.}(2002){Boldyrev}, {Nordlund}, \&
  {Padoan}}]{Boldyrev2002b}
{Boldyrev}, S., {Nordlund}, {\r{A}}., \& {Padoan}, P. 2002, \apj, 573, 678,
  \dodoi{10.1086/340758}

\bibitem[{{Brunt} {et~al.}(2009){Brunt}, {Heyer}, \& {Mac Low}}]{Brunt2009}
{Brunt}, C.~M., {Heyer}, M.~H., \& {Mac Low}, M.~M. 2009, \aap, 504, 883,
  \dodoi{10.1051/0004-6361/200911797}

\bibitem[{{Burkhart}(2021)}]{Burkhart2021}
{Burkhart}, B. 2021, \pasp, 133, 102001, \dodoi{10.1088/1538-3873/ac25cf}

\bibitem[{{Cai} {et~al.}(2021){Cai}, {Yang}, {Zheng}, {Yan}, {Zhang}, {Zhou},
  \& {Feng}}]{Cai2021}
{Cai}, J.-J., {Yang}, J., {Zheng}, S., {et~al.} 2021, Research in Astronomy and
  Astrophysics, 21, 304, \dodoi{10.1088/1674-4527/21/12/304}

\bibitem[{{Chira} {et~al.}(2019){Chira}, {Ib{\'a}{\~n}ez-Mej{\'\i}a}, {Mac
  Low}, \& {Henning}}]{Chira2019}
{Chira}, R.~A., {Ib{\'a}{\~n}ez-Mej{\'\i}a}, J.~C., {Mac Low}, M.~M., \&
  {Henning}, T. 2019, \aap, 630, A97, \dodoi{10.1051/0004-6361/201833970}

\bibitem[{{DeMarco} \& {Basu}(2017)}]{DeMarco2017}
{DeMarco}, A.~W., \& {Basu}, S. 2017, \pre, 95, 052114,
  \dodoi{10.1103/PhysRevE.95.052114}

\bibitem[{{Dudok de Wit}(2004)}]{Wit2004}
{Dudok de Wit}, T. 2004, \pre, 70, 055302, \dodoi{10.1103/PhysRevE.70.055302}

\bibitem[{Ester {et~al.}(1996)Ester, Kriegel, Sander, Xu, {et~al.}}]{Ester1996}
Ester, M., Kriegel, H.-P., Sander, J., Xu, X., {et~al.} 1996, in Kdd, Vol.~96,
  226--231

\bibitem[{{Federrath} {et~al.}(2010){Federrath}, {Roman-Duval}, {Klessen},
  {Schmidt}, \& {Mac Low}}]{Federrath2010}
{Federrath}, C., {Roman-Duval}, J., {Klessen}, R.~S., {Schmidt}, W., \& {Mac
  Low}, M.~M. 2010, \aap, 512, A81, \dodoi{10.1051/0004-6361/200912437}

\bibitem[{Frisch(1995)}]{Frisch1995}
Frisch, U. 1995, Turbulence: The Legacy of A. N. Kolmogorov (Cambridge
  University Press), \dodoi{10.1017/CBO9781139170666}

\bibitem[{{Frisch} {et~al.}(1978){Frisch}, {Sulem}, \& {Nelkin}}]{Frisch1978}
{Frisch}, U., {Sulem}, P.~L., \& {Nelkin}, M. 1978, Journal of Fluid Mechanics,
  87, 719, \dodoi{10.1017/S0022112078001846}

\bibitem[{{Gill} \& {Henriksen}(1990)}]{Gill1990}
{Gill}, A.~G., \& {Henriksen}, R.~N. 1990, \apjl, 365, L27,
  \dodoi{10.1086/185880}

\bibitem[{{Goodman} {et~al.}(2009){Goodman}, {Pineda}, \&
  {Schnee}}]{Goodman2009}
{Goodman}, A.~A., {Pineda}, J.~E., \& {Schnee}, S.~L. 2009, \apj, 692, 91,
  \dodoi{10.1088/0004-637X/692/1/91}

\bibitem[{{Henshaw} {et~al.}(2020){Henshaw}, {Kruijssen}, {Longmore}, {Riener},
  {Leroy}, {Rosolowsky}, {Ginsburg}, {Battersby}, {Chevance}, {Meidt},
  {Glover}, {Hughes}, {Kainulainen}, {Klessen}, {Schinnerer}, {Schruba},
  {Beuther}, {Bigiel}, {Blanc}, {Emsellem}, {Henning}, {Herrera}, {Koch},
  {Pety}, {Ragan}, \& {Sun}}]{Henshaw2020}
{Henshaw}, J.~D., {Kruijssen}, J.~M.~D., {Longmore}, S.~N., {et~al.} 2020,
  Nature Astronomy, 4, 1064, \dodoi{10.1038/s41550-020-1126-z}

\bibitem[{{Heyer} \& {Brunt}(2004)}]{Heyer2004}
{Heyer}, M.~H., \& {Brunt}, C.~M. 2004, \apjl, 615, L45, \dodoi{10.1086/425978}

\bibitem[{{Heyer} \& {Peter Schloerb}(1997)}]{Heyer1997}
{Heyer}, M.~H., \& {Peter Schloerb}, F. 1997, \apj, 475, 173,
  \dodoi{10.1086/303514}

\bibitem[{{Hily-Blant} {et~al.}(2008){Hily-Blant}, {Falgarone}, \&
  {Pety}}]{Hily-Blant2008}
{Hily-Blant}, P., {Falgarone}, E., \& {Pety}, J. 2008, \aap, 481, 367,
  \dodoi{10.1051/0004-6361:20078423}

\bibitem[{{Hu} {et~al.}(2022){Hu}, {Federrath}, {Xu}, \& {Mathew}}]{Hu2022}
{Hu}, Y., {Federrath}, C., {Xu}, S., \& {Mathew}, S.~S. 2022, \mnras, 513,
  2100, \dodoi{10.1093/mnras/stac972}

\bibitem[{{Jim{\'e}nez}(1998)}]{Jimenez1998}
{Jim{\'e}nez}, J. 1998, European Journal of Mechanics, B/Fluids, 17, 405,
  \dodoi{10.1016/S0997-7546(98)80002-2}

\bibitem[{{Joung} \& {Mac Low}(2006)}]{Joung2006}
{Joung}, M.~K.~R., \& {Mac Low}, M.-M. 2006, \apj, 653, 1266,
  \dodoi{10.1086/508795}

\bibitem[{{Kolmogorov}(1941{\natexlab{a}})}]{Kolmogorov1941a}
{Kolmogorov}, A. 1941{\natexlab{a}}, Akademiia Nauk SSSR Doklady, 30, 301

\bibitem[{{Kolmogorov}(1941{\natexlab{b}})}]{Kolmogorov1941b}
{Kolmogorov}, A.~N. 1941{\natexlab{b}}, Akademiia Nauk SSSR Doklady, 32, 16

\bibitem[{{Kolmogorov}(1962)}]{Kolmogorov1962}
---. 1962, Journal of Fluid Mechanics, 13, 82,
  \dodoi{10.1017/S0022112062000518}

\bibitem[{{Kritsuk} {et~al.}(2007){Kritsuk}, {Norman}, {Padoan}, \&
  {Wagner}}]{Kritsuk2007}
{Kritsuk}, A.~G., {Norman}, M.~L., {Padoan}, P., \& {Wagner}, R. 2007, \apj,
  665, 416, \dodoi{10.1086/519443}

\bibitem[{{Li} {et~al.}(2018){Li}, {Wang}, {Zhang}, {Ma}, {Fang}, \&
  {Yang}}]{Li2018}
{Li}, C., {Wang}, H., {Zhang}, M., {et~al.} 2018, \apjs, 238, 10,
  \dodoi{10.3847/1538-4365/aad963}

\bibitem[{{Lis} {et~al.}(1996){Lis}, {Pety}, {Phillips}, \&
  {Falgarone}}]{Lis1996}
{Lis}, D.~C., {Pety}, J., {Phillips}, T.~G., \& {Falgarone}, E. 1996, \apj,
  463, 623, \dodoi{10.1086/177276}

\bibitem[{{Ma} {et~al.}(2021){Ma}, {Wang}, {Li}, {Lin}, {Sun}, \&
  {Yang}}]{Ma2021}
{Ma}, Y., {Wang}, H., {Li}, C., {et~al.} 2021, \apjs, 254, 3,
  \dodoi{10.3847/1538-4365/abe85c}

\bibitem[{{Ma} {et~al.}(2022){Ma}, {Wang}, {Zhang}, {Wang}, {Zhang}, {Liu},
  {Li}, {Zheng}, {Yuan}, \& {Yang}}]{Ma2022}
{Ma}, Y., {Wang}, H., {Zhang}, M., {et~al.} 2022, \apjs, 262, 16,
  \dodoi{10.3847/1538-4365/ac7797}

\bibitem[{Mac~Low \& Klessen(2004)}]{Mac2004}
Mac~Low, M.-M., \& Klessen, R.~S. 2004, Reviews of modern physics, 76, 125

\bibitem[{{Mac Low} {et~al.}(1998){Mac Low}, {Klessen}, {Burkert}, \&
  {Smith}}]{MacLow1998}
{Mac Low}, M.-M., {Klessen}, R.~S., {Burkert}, A., \& {Smith}, M.~D. 1998,
  \prl, 80, 2754, \dodoi{10.1103/PhysRevLett.80.2754}

\bibitem[{{MacLow}(2004)}]{MacLow2004}
{MacLow}, M.-M. 2004, \apss, 289, 323,
  \dodoi{10.1023/B:ASTR.0000014961.72318.7c}

\bibitem[{{Meneveau} \& {Sreenivasan}(1991)}]{Meneveau1991}
{Meneveau}, C., \& {Sreenivasan}, K.~R. 1991, Journal of Fluid Mechanics, 224,
  429, \dodoi{10.1017/S0022112091001830}

\bibitem[{Moeckel \& Burkert(2015)}]{Moeckel2015}
Moeckel, N., \& Burkert, A. 2015, The Astrophysical Journal, 807, 67,
  \dodoi{10.1088/0004-637X/807/1/67}

\bibitem[{{Ossenkopf} {et~al.}(2008{\natexlab{a}}){Ossenkopf}, {Krips}, \&
  {Stutzki}}]{Ossenkopf2008a}
{Ossenkopf}, V., {Krips}, M., \& {Stutzki}, J. 2008{\natexlab{a}}, \aap, 485,
  917, \dodoi{10.1051/0004-6361:20079106}

\bibitem[{{Ossenkopf} {et~al.}(2008{\natexlab{b}}){Ossenkopf}, {Krips}, \&
  {Stutzki}}]{Ossenkopf2008b}
---. 2008{\natexlab{b}}, \aap, 485, 719, \dodoi{10.1051/0004-6361:20079107}

\bibitem[{{Padoan} {et~al.}(2003){Padoan}, {Boldyrev}, {Langer}, \&
  {Nordlund}}]{Padoan2003}
{Padoan}, P., {Boldyrev}, S., {Langer}, W., \& {Nordlund}, {\r{A}}. 2003, \apj,
  583, 308, \dodoi{10.1086/345351}

\bibitem[{{Padoan} {et~al.}(2004){Padoan}, {Jimenez}, {Nordlund}, \&
  {Boldyrev}}]{Padoan2004}
{Padoan}, P., {Jimenez}, R., {Nordlund}, {\r{A}}., \& {Boldyrev}, S. 2004,
  \prl, 92, 191102, \dodoi{10.1103/PhysRevLett.92.191102}

\bibitem[{{Padoan} \& {Nordlund}(2011)}]{Padoan2011}
{Padoan}, P., \& {Nordlund}, {\r{A}}. 2011, \apj, 730, 40,
  \dodoi{10.1088/0004-637X/730/1/40}

\bibitem[{{Pety} \& {Falgarone}(2003)}]{Pety2003}
{Pety}, J., \& {Falgarone}, E. 2003, \aap, 412, 417,
  \dodoi{10.1051/0004-6361:20031474}

\bibitem[{Pope(2001)}]{Pope2001}
Pope, S.~B. 2001, Measurement Science and Technology, 12, 2020,
  \dodoi{10.1088/0957-0233/12/11/705}

\bibitem[{{Reid} {et~al.}(2019){Reid}, {Menten}, {Brunthaler}, {Zheng}, {Dame},
  {Xu}, {Li}, {Sakai}, {Wu}, {Immer}, {Zhang}, {Sanna}, {Moscadelli}, {Rygl},
  {Bartkiewicz}, {Hu}, {Quiroga-Nu{\~n}ez}, \& {van Langevelde}}]{Reid2019}
{Reid}, M.~J., {Menten}, K.~M., {Brunthaler}, A., {et~al.} 2019, \apj, 885,
  131, \dodoi{10.3847/1538-4357/ab4a11}

\bibitem[{Richardson(1922)}]{Richardson1922}
Richardson, L.~F. 1922, Weather prediction by numerical process (University
  Press)

\bibitem[{{Roman-Duval} {et~al.}(2011){Roman-Duval}, {Federrath}, {Brunt},
  {Heyer}, {Jackson}, \& {Klessen}}]{Roman-Duval2011}
{Roman-Duval}, J., {Federrath}, C., {Brunt}, C., {et~al.} 2011, \apj, 740, 120,
  \dodoi{10.1088/0004-637X/740/2/120}

\bibitem[{{Schertzer} \& {Lovejoy}(1987)}]{Schertzer1987}
{Schertzer}, D., \& {Lovejoy}, S. 1987, \jgr, 92, 9693,
  \dodoi{10.1029/JD092iD08p09693}

\bibitem[{{Schmidt} {et~al.}(2009){Schmidt}, {Federrath}, {Hupp}, {Kern}, \&
  {Niemeyer}}]{Schmidt2009}
{Schmidt}, W., {Federrath}, C., {Hupp}, M., {Kern}, S., \& {Niemeyer}, J.~C.
  2009, \aap, 494, 127, \dodoi{10.1051/0004-6361:200809967}

\bibitem[{{Shan} {et~al.}(2012){Shan}, {Yang}, {Shi}, {Yao}, {Zuo}, {Lin},
  {Chen}, {Zhang}, {Duan}, {Cao}, {Li}, {Li}, {Liu}, \& {Zhong}}]{Shan2012}
{Shan}, W., {Yang}, J., {Shi}, S., {et~al.} 2012, IEEE Transactions on
  Terahertz Science and Technology, 2, 593, \dodoi{10.1109/TTHZ.2012.2213818}

\bibitem[{{She} \& {Leveque}(1994)}]{She1994}
{She}, Z.-S., \& {Leveque}, E. 1994, \prl, 72, 336,
  \dodoi{10.1103/PhysRevLett.72.336}

\bibitem[{{She} {et~al.}(2001){She}, {Ren}, {Lewis}, \& {Swinney}}]{She2001}
{She}, Z.-S., {Ren}, K., {Lewis}, G.~S., \& {Swinney}, H.~L. 2001, \pre, 64,
  016308, \dodoi{10.1103/PhysRevE.64.016308}

\bibitem[{{She} \& {Waymire}(1995)}]{She1995}
{She}, Z.-S., \& {Waymire}, E.~C. 1995, \prl, 74, 262,
  \dodoi{10.1103/PhysRevLett.74.262}

\bibitem[{{She} \& {Zhang}(2009)}]{She2009}
{She}, Z.-S., \& {Zhang}, Z.-X. 2009, Acta Mechanica Sinica, 25, 279,
  \dodoi{10.1007/s10409-009-0257-3}

\bibitem[{{Stewart} \& {Federrath}(2022)}]{Stewart2022}
{Stewart}, M., \& {Federrath}, C. 2022, \mnras, 509, 5237,
  \dodoi{10.1093/mnras/stab3313}

\bibitem[{{Su} {et~al.}(2019){Su}, {Yang}, {Zhang}, {Gong}, {Wang}, {Zhou},
  {Wang}, {Chen}, {Sun}, {Chen}, {Xu}, \& {Jiang}}]{Su2019}
{Su}, Y., {Yang}, J., {Zhang}, S., {et~al.} 2019, \apjs, 240, 9,
  \dodoi{10.3847/1538-4365/aaf1c8}

\bibitem[{{Vazquez-Semadeni}(1994)}]{Vazquez-Semadeni1994}
{Vazquez-Semadeni}, E. 1994, \apj, 423, 681, \dodoi{10.1086/173847}

\bibitem[{{V{\'a}zquez-Semadeni} {et~al.}(2003){V{\'a}zquez-Semadeni},
  {Ballesteros-Paredes}, \& {Klessen}}]{Vazquez-Semadeni2003}
{V{\'a}zquez-Semadeni}, E., {Ballesteros-Paredes}, J., \& {Klessen}, R.~S.
  2003, \apjl, 585, L131, \dodoi{10.1086/374325}

\bibitem[{{V{\'a}zquez-Semadeni} {et~al.}(2019){V{\'a}zquez-Semadeni}, {Palau},
  {Ballesteros-Paredes}, {G{\'o}mez}, \&
  {Zamora-Avil{\'e}s}}]{Vazquez-Semadeni2019}
{V{\'a}zquez-Semadeni}, E., {Palau}, A., {Ballesteros-Paredes}, J.,
  {G{\'o}mez}, G.~C., \& {Zamora-Avil{\'e}s}, M. 2019, \mnras, 490, 3061,
  \dodoi{10.1093/mnras/stz2736}

\bibitem[{{V{\'a}zquez-Semadeni} {et~al.}(2024){V{\'a}zquez-Semadeni}, {Palau},
  {G{\'o}mez}, {Arroyo-Ch{\'a}vez}, {Alig}, {Ballesteros-Paredes}, {Camacho},
  {Gonz{\'a}lez-Samaniego}, \& {Burkert}}]{Vazquez-Semadeni2024}
{V{\'a}zquez-Semadeni}, E., {Palau}, A., {G{\'o}mez}, G.~C., {et~al.} 2024,
  arXiv e-prints, arXiv:2408.10406, \dodoi{10.48550/arXiv.2408.10406}

\bibitem[{{Vincent} \& {Meneguzzi}(1991)}]{Vincen1991}
{Vincent}, A., \& {Meneguzzi}, M. 1991, Journal of Fluid Mechanics, 225, 1,
  \dodoi{10.1017/S0022112091001957}

\bibitem[{{Wenger} {et~al.}(2018){Wenger}, {Balser}, {Anderson}, \&
  {Bania}}]{Wenger2018}
{Wenger}, T.~V., {Balser}, D.~S., {Anderson}, L.~D., \& {Bania}, T.~M. 2018,
  \apj, 856, 52, \dodoi{10.3847/1538-4357/aaaec8}

\bibitem[{{Yan} {et~al.}(2020){Yan}, {Yang}, {Su}, {Sun}, \& {Wang}}]{Yan2020}
{Yan}, Q.-Z., {Yang}, J., {Su}, Y., {Sun}, Y., \& {Wang}, C. 2020, \apj, 898,
  80, \dodoi{10.3847/1538-4357/ab9f9c}

\bibitem[{{Yan} {et~al.}(2021){Yan}, {Yang}, {Su}, {Sun}, {Xu}, {Wang}, {Zhou},
  \& {Wang}}]{Yan2021}
{Yan}, Q.-Z., {Yang}, J., {Su}, Y., {et~al.} 2021, \apj, 922, 8,
  \dodoi{10.3847/1538-4357/ac214f}

\end{thebibliography}
\bibliographystyle{aasjournal}


% \listofchanges

\end{document}